\documentclass[a4paper,11pt]{article}
\pdfoutput=1

\usepackage{hyperref}

\usepackage{
graphicx,
hyperref,
amsmath,
amssymb,
charter,
xcolor,
ifluatex,
multirow}
        
\definecolor{Gray}{gray}{0.95}
\definecolor{RGray}{gray}{0.85}
\definecolor{CGray}{gray}{0.92}

\definecolor{tit}{rgb}{0.1,0.2,0.4}
\definecolor{blus}{cmyk}{1,1,0,0.6}
\definecolor{verde}{cmyk}{0.92,0,0.59,0.25}
        
\newcommand{\e}[1]{\cdot 10^{#1}}
\usepackage{tipa}
\usepackage{amsmath,amssymb,amsfonts, bm}
\usepackage{epsfig}
\usepackage{graphicx}
\usepackage{slashed}
\usepackage{color}

\usepackage{caption}
\usepackage{subcaption}
\usepackage{slashed} 
\usepackage{float}
\usepackage{comment}

\usepackage{booktabs}
\usepackage[normalem]{ulem}

\newcommand{\D}{{\cal D}}
\newcommand{\U}{{\cal U}}

\newcommand{\M}{{\cal M}}

\newcommand{\bbbar}{{b \bar{b}}}

\usepackage{calc}

\newcommand{\be}{\begin{equation}}
\newcommand{\ee}{\end{equation}}

\newcommand{\bea}{\begin{eqnarray}}
\newcommand{\eea}{\end{eqnarray}}

\newcommand{\bfig}{\begin{figure}}
\newcommand{\efig}{\end{figure}}

\newcommand{\lag}{\ensuremath{\mathcal{L}}}

\newcommand{\br}{\text{BR}}

\newcommand*{\myalign}[2]{\multicolumn{1}{#1}{#2}}

\newcommand{\iab}{{\ensuremath\rm ab^{-1}}}

\makeatletter
\newcommand*{\rom}[1]{\expandafter\@slowromancap\romannumeral #1@}
\makeatother

\begin{document}
\allowdisplaybreaks
\vspace*{-2.5cm}
\begin{flushright}
{\small
IIT-BHU
}
\end{flushright}

\vspace{2cm}

\begin{center}
{\LARGE \bf \color{tit} Phenomenology of the standard HVM and 95.4 GeV excess }\\[1cm]

{\large\bf Gauhar Abbas$^{a}$\footnote{email: gauhar.phy@iitbhu.ac.in} \\  [7mm]
Neelam  Singh$^{a}$\footnote{email: neelamsingh.rs.phy19@itbhu.ac.in }   } \\[7mm]
{\it $^a$ } {\em Department of Physics, Indian Institute of Technology (BHU), Varanasi 221005, India}\\[3mm]

\vspace{1cm}
{\large\bf\color{blus} Abstract}
\begin{quote}
We investigate the collider phenomenology of the standard Hierarchical VEVs Model by proposing a new version, which avoids large flavor changing neutral current interactions, thus, rendering the scale of new physics as low as the electroweak scale.  The resulting collider signatures are distinctive and testable at the High-Luminosity LHC, the High-Energy LHC, and future 100\,TeV hadron colliders. Remarkably, one of the pseudoscalars in the model can account for the 95.4\,GeV di-photon excess observed by ATLAS and CMS. In addition, the model naturally accommodates a new class of neutrino-philic dark matter candidate, \emph{neutrinic dark matter}, that interacts exclusively with neutrino pairs.

\end{quote}
\thispagestyle{empty}
\end{center}

\begin{quote}
{\large\noindent\color{blus} 
}

\end{quote}

\newpage
\setcounter{footnote}{0}
\section{Introduction}

The origin of matter in the Standard Model (SM) presents a profound and multifaceted theoretical challenge. This problem encompasses two major components: the existence of a hierarchical fermion mass spectrum and fermion mixing, including in the lepton sector, and confirmed the presence of dark matter in the Universe. The former aspect is commonly referred to as the flavor problem of the SM~\cite{Abbas:2023ivi,Abbas:2024wzp}. Various solutions to this problem have been proposed over the years, including mechanisms based on flavor symmetries, radiative mass generation, and compositeness~\cite{Froggatt:1978nt}–\cite{Chatterjee:2024dgw}.

The flavor problem can be addressed by introducing hierarchical vacuum expectation values (VEVs) of gauge-singlet scalar fields, which effectively distinguish between SM fermions both among and within generations~\cite{Abbas:2017vws,Abbas:2020frs,Abbas:2023dpf}. This framework is known as the Hierarchical VEV Model (HVM)~\cite{Abbas:2017vws}.  A simple realization of the HVM arises through dimension-five operators of the form:
\begin{equation}
y_{ij}~ \bar{\psi}_{L_i} \varphi \psi_{R_j} \frac{\chi_r}{\Lambda},
\end{equation}
where $y_{ij}$ are dimensionless couplings, $\psi_{L_i}$ denote the SM left-handed fermion doublets, $\psi_{R_j}$ are the SM right-handed singlet fermions, $\varphi$ is the Higgs doublet, and $\Lambda$ represents the scale at which the operator is generated and renormalized.

The HVM approach is built on the idea that the gauge-singlet scalar fields $\chi_r$ are not fundamental scalars. Instead, they are interpreted as multi-fermion bound states arising from a strongly coupled dark-technicolor gauge dynamics, as discussed in the appendix. A new realization of this framework is the Standard Hierarchical VEV Model (SHVM), which has been shown to make precise predictions for leptonic mixing parameters in terms of the Cabibbo angle and the masses of the strange ($s$) and charm ($c$) quarks~\cite{Abbas:2023dpf}.

The SHVM can naturally be embedded within the dark-technicolor (DTC) paradigm, where the hierarchical VEVs emerge as multi-fermion chiral condensates~\cite{Abbas:2020frs,Abbas:2025ser}. As shown in reference~\cite{Abbas:2025ser}, the DTC paradigm is based on QCD-like gauge dynamics, which can generate electroweak symmetry breaking while remaining consistent with electroweak precision observables. Furthermore, the DTC paradigm can address the flavor problem either through the SHVM or the Froggatt-Nielsen mechanism~\cite{Froggatt:1978nt}.

Ultraviolet (UV) completion of the SHVM, the DTC paradigm, features three distinct sectors of strong gauge dynamics: $SU(N)_{\rm TC}$, $SU(N)_{\rm DTC}$, and $SU(N)_{\rm F}$. Here, TC denotes the technicolor sector responsible for electroweak symmetry breaking, DTC corresponds to the dark-technicolor dynamics underlying the hierarchical VEV structure, and F refers to a dark-QCD–like sector composed of vector-like fermions. 

The SHVM may offer an intriguing framework in light of several recently reported anomalies spanning mass scales from around 10\,GeV to the TeV range. These anomalies, summarized in a recent review by the ATLAS collaboration~\cite{ATLAS:2024itc}, are listed in Table~\ref{tab:anomalies}. Notably, the corresponding mass values appear both hierarchical and seemingly random, posing a significant challenge for conventional models based on extended scalar sectors.  If future runs of the Large Hadron Collider (LHC) continue to provide evidence supporting some of these anomalies, the SHVM, when embedded within the strongly coupled dynamics of the DTC paradigm, may emerge as a promising candidate to account for them. This possibility will be partially explored in the present work.

\begin{table}[H]
\begin{center}
\resizebox{0.6 \textwidth}{!}{
\begin{tabular}{l|c|c|}
\multirow{2}{*}{\textbf{Decay channel}} &	 \textbf{Production}	&	\multirow{2}{*}{\textbf{Mass GeV}}  \\
& \textbf{mode}					&	\\ \hline
$H\to\tau\tau$ &											$b$-associated &		400 	\\			
$H\to\tau\tau$ &											ggF &								400 		 \\
$H\to\mu\mu$ &												$b$-associated &		480 \\
$H\to t\bar{t}$ &											ggF &								800 	 \\
$H\to t\bar{t}/t\bar{q}$ &						qq and qg &					900  \\
$H\to ZZ\to 4\ell /2\ell 2\nu$ &			ggF &								240  \\
$H\to ZZ\to 4\ell /2\ell 2\nu$ &			VBF &								620  \\
$H\to \gamma\gamma$ &									ggF &								684 	 \\
$H\to \gamma\gamma$ &									ggF &								~~~~~95.4  \\
$H\to Z(\ell\ell)\gamma$ &						ggF &								420	 \\
$H\to Z(q\bar{q})\gamma$ &						ggF &								3640 \\
$A\to Zh_{125}(b\bar{b})$ &						ggF &								500  \\
$A\to Zh_{125}(b\bar{b})$ &						$b$-associated &		500  \\
$A\to ZH\to \ell\ell b\bar{b}$ &			ggF	&								610 ($A$), 290 ($H$)  \\
$A\to ZH\to \ell\ell b\bar{b}$ &			$b$-associated &		440 ($A$), 220 ($H$) \\
$A\to ZH\to \ell\ell WW$ &						ggF	&								440 ($A$), 310 ($H$)  \\
$A\to ZH\to \ell\ell t\bar{t}$ &			ggF	&								650 ($A$), 450 ($H$)  \\
$A\to ZH\to Zh_{125}(b\bar{b})h_{125}(b\bar{b})$ &	VH &	420 ($A$), 320 ($H$)   \\
$H^+\to cb$ &													$t\bar{t}$ decay &	130  \\
$H^+\to Wa(\mu\mu )$ &								$t\bar{t}$ decay &	120--160 ($H^+$), 27 ($a$)  \\
$H^{++}\to WW$ &											VBF &								450 	 \\
$H\to h_{125}h_{125}\to 4b$ &					ggF &								1100 \\
$H\to h_{125}h_{125}\to 4b$ &					VBF &								550 	 \\
$H\to h_{125}h_{125}\to b\bar{b}\tau\tau$ &	ggF &					1000~~  \\
$H\to h_{125}h_{125} \, \text{combination}$ &	ggF &				1100~~ \\
$X\to Sh_{125}\to \bbbar\gamma\gamma$ &	ggF &							575 ($X$), 200 ($S$)  \\
$h_{125}\to Z_dZ_d \to 4\ell$	&				ggF &								~~28  \\
$h_{125}\to ZZ_d \to 4\ell$ &					ggF &								~~39 \\
$h_{125}\to aa\to b\bar{b}\mu\mu$ &		ggF, VBF, VH &			~~52 \\
$h_{125}\to aa\to 4\gamma$ &					ggF &								10--25~~~~  \\
$h_{125}\to e\tau$ and $h_{125}\to \mu\tau$ &		ggF, VBF, VH &	125~~ 
 \end{tabular}
}
\caption{  The anomalies reported by the ATLAS experiment. Here, $H$, $X$, and $S$ denote generic neutral scalars beyond the Standard Model (SM), while $a$ and $A$ represent beyond-the-SM pseudoscalars. The symbols $H^+$ and $H^{++}$ refer to generic singly and doubly charged Higgs bosons, respectively, and $h_{125}$ denotes the SM Higgs boson.  $b$ denotes the bottom quark, and $Z_d$ stands for a dark-vector boson. The third column specifies the mass hypotheses at which the corresponding beyond-the-SM scalars and pseudoscalars are searched for.  This table is adapted from reference \cite{ATLAS:2024itc}.}
\label{tab:anomalies}
\end{center}
\end{table}

In this work, we explore a scenario in which the $\mathcal{Z}_{N} \times \mathcal{Z}_{M} \times \mathcal{Z}_{P}$ flavor symmetry of the SHVM is softly broken. In this context, we introduce a new protection mechanism against large flavor-changing neutral currents (FCNCs), enabling the new physics scale $\Lambda$ to be lowered to the electroweak regime. This setup allows for a rich and testable collider phenomenology at the High-Luminosity LHC (HL-LHC), the High-Energy LHC (HE-LHC)~\cite{FCC:2018bvk}, and a future 100\,TeV high-luminosity hadron collider such as FCC-hh~\cite{FCC:2018byv}. In particular, we identify a set of non-overlapping and mutually exclusive collider signatures unique to the SHVM. These fingerprint signatures set the SHVM apart from all existing models with extended scalar sectors.

Furthermore, the axial degree of freedom of the gauge-singlet scalar field $\chi_7$ gives rise to a novel class of neutrino-philic dark matter, which we refer to as \textit{neutrinic dark matter}. A defining feature of this candidate is its absence of coupling to the electromagnetic field, distinguishing it sharply from other ALP dark matter scenarios. This makes neutrinic dark matter a unique and minimal possibility within the SHVM framework.

This paper is organized as follows: in Section~\ref{SHVM}, we propose a new version of the  SHVM. The scalar potential and the protection mechanism of SHVM against large FCNCs are discussed in Section~\ref{scalar_potential}. Collider signatures are discussed in Section~\ref{collider}, and the 95.4 GeV anomaly from the ATLAS review~\cite{ATLAS:2024itc} is considered in Section~\ref{anomaly}. Finally, neutrinic dark matter is discussed in Section~\ref{neutrinic_dm}, and we summarize our findings in Section~\ref{sum}.

\section{The standard HVM}
\label{SHVM}
In this section, we propose a new version of  the SHVM which can bypass the bounds from the large FCNC processes.  The SHVM is constructed by imposing a generic   $\mathcal{Z}_{\rm N} \times \mathcal{Z}_{\rm M} \times \mathcal{Z}_{\rm P}$ flavor symmetry on the SM, and on the gauge singlet scalar fields $\chi_r$.  The  fields $\chi_r$  are assumed to be the bound states of  strong dark-technicolor dynamics \cite{Abbas:2017vws,Abbas:2020frs,Abbas:2023dpf,Abbas:2025ser}.

The  $\mathcal{Z}_{\rm N} \times \mathcal{Z}_{\rm M} \times \mathcal{Z}_{\rm P}$ flavor symmetry naturally emerges in the DTC paradigm  through the breaking of three axial $U(1)_{A}$ symmetries  \cite{Abbas:2020frs,Abbas:2025ser}.  The symmetries $\rm SU(\rm N)_{\mathrm{TC,DTC,DQCD}}$ in the DTC paradigm are associated with three  global axial $U(1)_A^{\rm TC, DTC,  F}$ symmetries.  The instanton interactions of the corresponding QCD-like strong dynamics break the global axial $U(1)_A^{\rm TC, DTC,  F}$ symmetries as ~ \cite{Harari:1981bs},

\begin{align}
\mathrm{U(1)}_{X_{\mathrm{TC,DTC,DQCD}}} \longrightarrow  \mathbb{Z}_{2K_{\mathrm{TC,DTC,DQCD}}}.
\end{align}
where $\rm K_{\mathrm{TC,DTC,DQCD}}$ represents the  massless flavors of the $\rm SU(\rm N)_{\mathrm{TC,DTC,DQCD}}$  gauge symmetries in the $\rm N$-dimensional representation.  This results in a generic $\mathcal{Z}_{\rm N} \times \mathcal{Z}_{\rm M} \times \mathcal{Z}_{\rm P}$ flavor symmetry, where $\rm N= 2 \rm K_{\rm TC}$, $\rm M= 2 \rm K_{\rm DTC}$ and $\rm P= 2 \rm K_{\rm F}$ \cite{Harari:1981bs}. See appendix for more details.

The transformations of the composite scalar fields  $\chi _r $  under the $\mathcal{G}_{\rm SM} \equiv SU(3)_c \otimes SU(2)_L \otimes U(1)_Y$ symmetry of the SM are given as,
\begin{eqnarray}
\chi _r :(1,1,0),
 \end{eqnarray} 
where $r=1-6$.

After imposing the $\mathcal{Z}_{\rm N} \times \mathcal{Z}_{\rm M} \times \mathcal{Z}_{\rm P}$ flavor symmetry on the SM and the scalar fields  $\chi _r $, we can recover the masses of the charged fermions through  the  dimension-5 operators as given below \cite{Abbas:2017vws,Abbas:2023dpf},
\bea
\label{mass2}
{\mathcal{L}} &=& \dfrac{1}{\Lambda }\Bigl[  y_{ij}^u  \bar{\psi}_{L_i}^{q}  \tilde{\varphi} \psi_{R_j}^{u}   \chi _r +     
   y_{ij}^d  \bar{\psi}_{L_i}^{q}   \varphi \psi_{R_j}^{d}  \chi _{r}   +   y_{ij}^\ell  \bar{\psi}_{L_i}^{\ell}   \varphi \psi_{R_j}^{\ell}  \chi _{r} \Bigr]  
+  {\rm H.c.}.
\eea
where  $i$ and $j$   denote the family  indices, $ \psi_{L}^q,  \psi_{L}^\ell  $ are the quark and lepton left-handed doublets, $ \psi_{R}^u,  \psi_{R}^d, \psi_{R}^\ell$ show the right-handed up, down type  quarks and  leptons,  $\varphi$ and $ \tilde{\varphi}= -i \sigma_2 \varphi^* $  show the SM Higgs field and its conjugate,  and $\sigma_2$ stands for  the second Pauli matrix. 

In this work, we create a new version of the SHVM, which can avoid large FCNC interactions, and the new physics scale can be as low as 100 GeV.  This is achieved by assigning the following generic charges to the scalar and fermions  under the   $\mathcal{Z}_{\rm N} \times \mathcal{Z}_{\rm M} \times \mathcal{Z}_{\rm P}$ flavor symmetry,
\begin{align}
\psi_{L_1}^{q} &: (+, 1, \omega^{\rm P-3
}_{14}),~ \psi_{L_2}^{q}: (+, 1, \omega^{9}_{14}),~  \psi_{L_3}^{q}: (-, 1, \omega^{ 8}_{14}), \\ \nonumber
u_{R} &: (-, \omega_4, \omega^{\rm P-5}_{14}),~ c_{R}: (+, 1, \omega^4_{14}),~  t_{R}: (+, 1, 1), \\ \nonumber
d_{R} &: (-, \omega_4, \omega^{\rm 9}_{14}),~ s_{R}: (+, 1, \omega^{12}_{14}),~  b_{R}: (-, 1, \omega_{14}^2), \\ \nonumber
\psi_{L_1}^{\ell} &: (+, \omega^{3}_{4}, \omega^{12}_{14}),~ \psi_{L_2}^{\ell}: (+, \omega^{3}_4, \omega^{10}_{14}),~  \psi_{L_3}^{\ell}: (+, \omega^{3}_4, \omega^{6}_{14}), \\ \nonumber
e_{R} &: (-, 1, \omega^{10}_{14}),~ \mu_{R}: (+, \omega^{3}_4, \omega^{13}_{14}),~  \tau_{R}: (+, \omega^{3}_4, \omega_{14}), \\ \nonumber
\nu_{e_{R}} &: (+, 1, \omega^{8}_{14}),~ \nu_{\mu_{R}}: (-, \omega_4, \omega^{3}_{14}),~  \nu_{\tau_{R}}: (-, \omega_4, \omega_{14}^3), \\ \nonumber
\chi_1 &: (-, \omega^3_4, \omega^{2}_{14}),~ \chi_2: (+, 1, \omega^{5}_{14}),~  \chi_3: (-, 1, \omega^8_{14}), \\ \nonumber
\chi_4 &: (+, 1, \omega^{13}_{14}),~ \chi_5: (+, 1, \omega^{11}_{14}),~  \chi_6: (+, 1, \omega^{6}_{14}), 
\end{align}
where  $\omega_4$   and  $\omega_{14}$ represent  the fourth and  the fourteenth root of unity corresponding to the symmetries $\mathcal{Z}_4$  and $\mathcal{Z}_{14}$,  respectively, and   $\rm N= 2$, $\rm M\geq 4$,  and $\rm P \geq 14$.  For recovering the desired flavor structure of the charged fermions in the standard HVM,  we must have  $\rm N= 2$.

We obtain the neutrino masses and mixing by adding   three right-handed  neutrinos $\nu_{eR}$, $\nu_{\mu R}$, $\nu_{\tau R}$  to the SM through the following  dimension-6 Lagrangian,
\begin{eqnarray}
-{\mathcal{L}}_{\rm Yukawa}^{\nu} &=&      y_{ij}^\nu \bar{ \psi}_{L_i}^\ell   \tilde{\varphi}  \nu_{j_R} \left[  \dfrac{ \chi_r \chi_7 }{\Lambda^2} \right] +  {\rm H.c.},
\label{mass_Nu}
\end{eqnarray}
where $\chi_7$ is a gauge singlet scalar field, which is a bound state of a strong and dark QCD-like dynamics of vector-like fermions.  This is elaborated in the appendix.

The equation \ref{mass_Nu} places a severe constraint on the symmetry  $\mathcal{Z}_{\rm P}$ allowing only $\rm P = 14$, making it  a magic number. We show the  charge assignments of the fields for the normal ordering of neutrino masses as given in table   \ref{tab1} under the $\mathcal{Z}_2 \times \mathcal{Z}_{\rm 4} \times \mathcal{Z}_{14} $ flavor symmetry.  
 \begin{table}[ht]
\begin{center}
\begin{tabular}{|c|c|c|c||c|c|c|c||c|c|c|c||c|c|c|c|}
  \hline
  Fields                               &   $\mathcal{Z}_2$  &  $\mathcal{Z}_4$   &  $\mathcal{Z}_{14}$   & Fields   &  $\mathcal{Z}_2$   &  $\mathcal{Z}_4$ &  $\mathcal{Z}_{14}$ & Fields   & $\mathcal{Z}_2$  & $\mathcal{Z}_4$  &  $\mathcal{Z}_{14}$  & Fields  &  $\mathcal{Z}_2$   &  $\mathcal{Z}_4$  &  $\mathcal{Z}_{14}$ \\
  \hline
 $u_{R}$                        &     -   &     $\omega_4$          &    $\omega^{9}_{14}$ & $d_{R}$    &     -    & $\omega_4$        &    $\omega^{9}_{14}$  & $ \psi_{L_3}^{q} $       &    -     &  $1$    &   $\omega^8_{14}$   &  $\tau_R$      &   +  &  $\omega^3_{4}$      &     $\omega_{14}$             \\
  $c_{R}$                       &     +   &    $1$          &    $\omega^4_{14}$   &  $s_R$                         &      +  &  $1$      &  $\omega^{12}_{14}$  &  $ \psi_{L_1}^\ell $                          &     +   & $\omega^3_4$     &  $\omega^{12}_{14}$                          &   $\nu_{e_R}$   &    +   &   $1$     &      $\omega^{8}_{14}$        \\
   $t_{R}$                        &     +   &    $1$         &    $1$   & $b_R$                         &      -  &  $1$    &  $\omega^{2}_{14}$  &  $ \psi_{L_2}^{\ell} $     &      +  & $\omega^3_4$      &  $\omega^{10}_{14}$     & $\nu_{\mu_R}$                   &     -  &   $\omega_4$    &   $\omega^3_{14}$                    \\
  $\chi _1$                        &      -  &   $\omega^3_4$      &    $\omega^2_{14}$    &   $\chi _4$                          &      +  &  $1$      &   $ \omega^{13}_{14}$    &   $ \psi_{L_3}^{ \ell} $       &    +     &  $\omega^3_4$    &   $\omega^6_{14}$                                 & $\nu_{\tau_R}$                    &     -   &  $\omega_4$     &   $\omega^3_{14}$          \\
  $\chi _2$                   & +     &       $1$      &  $\omega^5_{14}$   & $ \psi_{L_1}^q $                          &      +  &  $1$      &  $\omega^{11}_{14}$  & $e_R$    &      -   &   $1$       &    $\omega^{10}_{14}$      &  $\chi_7 $                          &      -   &  $\omega^2_4$   &     $\omega^8_{14}$                                               \\
   $\chi _3$                  &    -   &       $1$    & $ \omega^8_{14}$        & $ \psi_{L_2}^{q} $     &      +  & $1 $      &  $\omega^{9}_{14}$  &   $ \mu_R$     &   +  & $\omega^3_4$       &     $\omega^{13}_{14}$      &  $ \varphi $                           &      +  &1     &   1                                            \\
    $\chi _5$                  &    +  &       $1$    & $ \omega^{11}_{14}$        &  $\chi _6$         &      +  & $1 $      &  $\omega^{6}_{14}$  &       &     &       &         &                           &       &     &                                             \\
  \hline
\end{tabular}
\end{center}
\caption{The charges of left- and right-handed fermions  and  scalar fields under $\mathcal{Z}_2$, $\mathcal{Z}_4$  and $\mathcal{Z}_{14}$ symmetries for the normal mass ordering. $\omega_4$  and  $\omega_{14}$ are the fourth and  the fourteenth root of unity corresponding to the symmetries $\mathcal{Z}_4$  and $\mathcal{Z}_{14}$,  respectively.}
 \label{tab1}
\end{table} 

The flavor problem of the SM can be addressed in terms of the     VEVs pattern    $ \langle \chi _4 \rangle > \langle \chi _1 \rangle $, $ \langle \chi _2 \rangle >> \langle \chi _5 \rangle $, $ \langle \chi _3 \rangle >> \langle \chi _6 \rangle $, $ \langle \chi _{3} \rangle >> \langle \chi _{2} \rangle >> \langle \chi _{1} \rangle $, and  $ \langle \chi _6 \rangle >> \langle \chi _5 \rangle >> \langle \chi _4 \rangle $.  The Lagrangian providing masses to charged fermions is now given as,
\begin{align}
\label{mass22}
{\mathcal{L}_{f}}  =& \dfrac{1}{\Lambda }\Bigl[  y_{11}^u  \bar{\psi}_{L_1}^{q}  \tilde{\varphi} \psi_{R_1}^{u}   \chi _1 +  y_{13}^u  \bar{\psi}_{L_1}^{q}  \tilde{\varphi} \psi_{R_3}^{u}   \chi _5  +  y_{22}^u  \bar{\psi}_{L_2}^{q}  \tilde{\varphi} \psi_{R_2}^{u}   \chi_2  +  y_{23}^u  \bar{\psi}_{L_2}^{q}  \tilde{\varphi} \psi_{R_3}^{u}   \chi_2^\dagger  
  +  y_{33}^u  \bar{\psi}_{L_3}^{q}  \tilde{\varphi} \psi_{R_3}^{u}   \chi_3 \\ \nonumber
  & +   y_{11}^d  \bar{\psi}_{L_1}^{q}   \varphi \psi_{R_1}^{d}  \chi_{1} +     
   y_{12}^d  \bar{\psi}_{L_1}^{q}   \varphi \psi_{R_2}^{d}  \chi_{4}           
    + y_{22}^d  \bar{\psi}_{L_2}^{q}   \varphi \psi_{R_2}^{d}  \chi_{5}  + y_{33}^d  \bar{\psi}_{L_3}^{q}   \varphi \psi_{R_3}^{d}  \chi_{6}  \\ \nonumber
   & +   y_{11}^\ell  \bar{\psi}_{L_1}^{\ell}   \varphi \psi_{R_1}^{\ell}  \chi _{1}  +   y_{12}^\ell  \bar{\psi}_{L_1}^{\ell}   \varphi \psi_{R_2}^{\ell}  \chi _{4}  +   y_{13}^\ell  \bar{\psi}_{L_1}^{\ell}   \varphi \psi_{R_3}^{\ell}  \chi _{5}  +   y_{22}^\ell  \bar{\psi}_{L_2}^{\ell}   \varphi \psi_{R_2}^{\ell}  \chi _{5}  +   y_{23}^\ell  \bar{\psi}_{L_2}^{\ell}   \varphi \psi_{R_3}^{\ell}  \chi _{2}^\dagger  \\ \nonumber
& +   y_{33}^\ell  \bar{\psi}_{L_3}^{\ell}   \varphi \psi_{R_3}^{\ell}  \chi _{2} +  {\rm H.c.} \Bigr].
\end{align}

The mass matrices of the charged fermions can be written as,
\begin{align}
\label{mUD}
\M_\U & =   \dfrac{ v }{\sqrt{2}} 
\begin{pmatrix}
y_{11}^u  \epsilon_1 &  0  & y_{13}^u  \epsilon_{5}    \\
0    & y_{22}^u \epsilon_{2}  &  y_{23}^u  \epsilon_{2}   \\
0   &  0    &  y_{33}^u  \epsilon_{3} 
\end{pmatrix},  
\M_\D = \dfrac{ v }{\sqrt{2}} 
 \begin{pmatrix}
  y_{11}^d \epsilon_{1} &    y_{12}^d \epsilon_{4} &  0 \\
0 &     y_{22}^d \epsilon_{5} &  0\\
  0 &    0  &   y_{33}^d \epsilon_{6}\\
\end{pmatrix},  
\M_\ell  =\dfrac{ v }{\sqrt{2}} 
  \begin{pmatrix}
  y_{11}^\ell \epsilon_1 &    y_{12}^\ell \epsilon_4  &   y_{13}^\ell \epsilon_5 \\
 0 &    y_{22}^\ell \epsilon_5 &   y_{23}^\ell \epsilon_2\\
   0  &    0  &   y_{33}^\ell \epsilon_2 \\
\end{pmatrix},
\end{align} 
where $\epsilon_r = \dfrac{\langle \chi _{r} \rangle }{\Lambda}$ and  $\epsilon_r<1$.

The Lagrangian generating neutrino masses in the normal ordering is,
\bea
\label{mass2c}
{\mathcal{L}_{f}} &=& \dfrac{1}{\Lambda^2 }\Bigl[   y_{11}^\nu  \bar{\psi}_{L_1}^{\ell}  \tilde{\varphi} \psi_{R_1}^{\nu}   \chi _1^\dagger \chi_7^\dagger +  y_{12}^\nu  \bar{\psi}_{L_1}^{\ell}  \tilde{\varphi} \psi_{R_2}^{\nu}   \chi _4^\dagger \chi_7 +  y_{13}^\nu  \bar{\psi}_{L_1}^{\ell}  \tilde{\varphi} \psi_{R_3}^{\nu}   \chi _4^\dagger \chi_7 +  y_{22}^\nu  \bar{\psi}_{L_2}^{\ell}  \tilde{\varphi} \psi_{R_2}^{\nu}   \chi _4 \chi_7 \\  \nonumber
&& +  y_{23}^\nu  \bar{\psi}_{L_2}^{\ell}  \tilde{\varphi} \psi_{R_3}^{\nu}   \chi _4 \chi_7 +  y_{32}^\nu  \bar{\psi}_{L_3}^{\ell}  \tilde{\varphi} \psi_{R_2}^{\nu}   \chi_5 \chi_7 +  y_{33}^\nu  \bar{\psi}_{L_3}^{\ell}  \tilde{\varphi} \psi_{R_3}^{\nu}   \chi_5 \chi_7 +  {\rm H.c.} \Bigr].
\eea

The Dirac mass matrix for neutrinos is given by,
\begin{equation}
\label{NM}
\M_{\D} = \dfrac{v}{\sqrt{2}}  
\begin{pmatrix}
y_{11}^\nu   \epsilon_1 \epsilon_{7}   &  y_{12}^\nu   \epsilon_{4} \epsilon_{7}  & y_{13}^\nu  \epsilon_{4}  \epsilon_{7} \\
0   & y_{22}^\nu  \epsilon_{4}  \epsilon_{7} &  y_{23}^\nu  \epsilon_{4}  \epsilon_{7} \\
0   &   y_{32}^\nu  \epsilon_{5}  \epsilon_{7}   &  y_{33}^\nu  \epsilon_{5}  \epsilon_{7}
\end{pmatrix},
\end{equation}
where $\epsilon_{7} = \frac{\langle \chi _{7} \rangle}{\Lambda} <1$.

The diagonalization of the  fermionic mass matrices is performed through the bi-unitary transformations given by,
\begin{align}
 U_{u,d,\ell,\nu}^\dagger   \M_{\U, \D,\ell,\nu}V_{u,d,\ell,\nu} = \M_{\U,\D,\ell,\nu}^{\rm dia}
\end{align}
The numerical form of the fermionic  masses matrices, and bi-unitary transformation matrices  are given in the appendix.

The masses of charged fermions  are,
\begin{eqnarray}
\label{mass1a}
m_t  &=& \ \left|y^u_{33} \right| \epsilon_{3} v/\sqrt{2}, ~
m_c  = \   \left|y^u_{22} \epsilon_{2} \right|  v /\sqrt{2} ,~
m_u  =  |y_{11}^u  |\,  \epsilon_1 v /\sqrt{2},\nonumber \\
m_b  &\approx& \ |y^d_{33}| \epsilon_{6} v/\sqrt{2}, 
m_s  \approx \   \left|y^d_{22}  \right| \epsilon_{5} v /\sqrt{2},
m_d  \approx  \left|y_{11}^d    \right|\,  \epsilon_{1} v /\sqrt{2},\nonumber \\
m_\tau  &\approx& \ |y^\ell_{33}| \epsilon_{2} v/\sqrt{2}, ~
m_\mu  \approx \   |y^\ell_{22} | \epsilon_{5} v /\sqrt{2} ,~
m_e  =  |y_{11}^\ell   |\,  \epsilon_1 v /\sqrt{2}.
\end {eqnarray}

Assuming all $|y_{ij}^{u,d}|$ order 1, the quark mixing angles turn out to be,
\begin{eqnarray}
\sin \theta_{12}  & \simeq&   \left|\frac{ y_{12}^d}{ y_{22}^d} \right| \frac{ \epsilon_{4}}{ \epsilon_{5}}, ~ 
\sin \theta_{23}  \simeq   \left|\frac{ y_{23}^u}{ y_{33}^u} \right| \frac{ \epsilon_{2}}{ \epsilon_{3}}, 
\sin \theta_{13}  \simeq    \left|\frac{ y_{13}^u}{ y_{33}^u} \right| \frac{ \epsilon_{5}}{ \epsilon_{3}}.
\end{eqnarray}

The following values of the parameters  $\epsilon_r$, in general, are used to produce  the fermion  masses and mixing parameters,
\begin{equation}
\label{epsi}
\epsilon_1 = 3.16 \times 10^{-6},~ \epsilon_2 = 0.0031,~ \epsilon_3 = 0.87,~\epsilon_4 = 0.000061,~\epsilon_5 = 0.000270,~\epsilon_6 = 0.0054,~\epsilon_7 = 7.18 \times 10^{-10}.  
\end{equation}
These values are obtained by fitting the the fermionic masses and mixing at $1$ TeV.  This is  discussed in the appendix of the paper.

The neutrino  masses can be written as,
\begin{eqnarray}
\label{mass11}
m_3  &\approx&  |y^\nu_{33}|  \epsilon_{5} \epsilon_{7} v/\sqrt{2}, 
m_2  \approx     \left|y^\nu_{22} - \dfrac{y_{23}^\nu  y_{32}^\nu}{y_{33}^\nu} \right|  \epsilon_{4} \epsilon_{7} v /\sqrt{2},
m_1  \approx  |y_{11}^\nu  |\,  \epsilon_1 \epsilon_{7} v /\sqrt{2}.
\end{eqnarray}

The resulting masses of neutrinos are,
\begin{align}
\label{neut_mass}
\{m_3,m_2,m_1\} = \{5.05 \times 10^{-2}, 8.67 \times 10^{-3}, 2.67 \times 10^{-4} \}\, \text{eV}
\end{align}
for  the numerical values of $y_{ij}^\nu$  given in the appendix.

The main achievement of the SHVM  is  the neutrino mixing angles.  Assuming all the couplings  $|y_{ij}^\nu|$ order 1, they can be written as \cite{Abbas:2023dpf},
\begin{eqnarray}
\sin \theta_{12}^\ell  &\simeq&  \left|-{y_{12}^\nu  \over y_{22}^\nu } +{y_{12}^\ell \epsilon_4  \over y_{22}^\ell  \epsilon_5}+  {y_{23}^{\ell *} y_{13}^\nu \epsilon_4  \over y_{33}^\ell y_{33}^\nu  \epsilon_5}   \right| \geq   \left|-{y_{12}^\nu  \over y_{22}^\nu }   \right| -   \left| {y_{12}^\ell   \over y_{22}^\ell  } +  {y_{23}^{\ell *} y_{13}^\nu   \over y_{33}^\ell y_{33}^\nu  }  \right|  {\epsilon_4  \over   \epsilon_5} \approx  1 - 2 \sin \theta_{12}, \\ \nonumber
\sin \theta_{23}^\ell  &\simeq&  \left|{y_{23}^\ell  \over y_{33}^\ell } - {y_{23}^\nu \epsilon_4  \over y_{33}^\nu  \epsilon_5} \right| \geq   \left|{y_{23}^\ell  \over y_{33}^\ell }   \right| -   \left| {y_{23}^\nu   \over y_{33}^\nu  }  \right|   { \epsilon_4  \over  \epsilon_5}\approx  1 -  \sin \theta_{12}, \\ \nonumber
\sin \theta_{13}^\ell    &\simeq& \left|    -{y_{13}^\nu \epsilon_4  \over y_{33}^\nu  \epsilon_5} + {y_{13}^\ell   \epsilon_5  \over y_{33}^\ell  \epsilon_2 }    \right| \geq  \left|    - {y_{13}^\nu   \over y_{33}^\nu  }  \right|  { \epsilon_4  \over   \epsilon_5}-  \left| {y_{13}^\ell     \over y_{33}^\ell   }  \right| { \epsilon_5  \over   \epsilon_2}  \approx \sin \theta_{12} - \frac{m_s}{m_c},
\end{eqnarray}
where $ m_s /m_c =      \epsilon_5  /   \epsilon_2    $.  Thus, in the SHVM, the leptonic mixing angles are predicted  in terms of the Cabibbo angle and masses of strange and charm quarks.

This results in very precise predictions of the leptonic mixing angles \cite{Abbas:2023dpf},
\begin{align}
 \sin \theta_{12}^\ell =  0.55 \pm 0.00134,~ \sin \theta_{23}^\ell =  0.775 \pm 0.00067,~  \sin \theta_{13}^\ell = 0.1413 - 0.1509, 
\end{align}
which may be probed by next generation neutrino experiments such as as DUNE, Hyper-Kamiokande, and JUNO \cite{Huber:2022lpm}.
 
\section{Scalar potential of the standard HVM}
\label{scalar_potential}
To construct the scalar potential of the SHVM, we impose an extra $\mathcal{Z}_2^\prime$ symmetry on the model. Under this new  $\mathcal{Z}_2^\prime$ symmetry, the right-handed  fermions transform as  $\psi_{R_{u,d,\ell,\nu}} : -$, the singlet scalar fields  as $\chi_r: -$,  where $r=1-6$, and the field $\chi_7$ as $\chi_7:+$ .  This  forbids all cubic terms in the scalar potential without affecting the flavor structure of the model, and simplifies the phenomenology of the SHVM.  Thus, we write 
\bea 
V &=& \mu^2 \varphi^\dagger \varphi + \lambda (\varphi^\dagger \varphi)^2  +   \mu_{\chi_1}^2 |\chi_1|^2   +   \mu_{\chi_2}^2 |\chi_2|^2   +   \mu_{\chi_3}^2 |\chi_3|^2    +   \mu_{\chi_4}^2|\chi_4|^2    +   \mu_{\chi_5}^2 |\chi_5|^2    +   \mu_{\chi_6}^2 |\chi_6|^2    
 \\ \nonumber  
 &&+   \mu_{\chi_7}^2 |\chi_7|^2 
+ \lambda_{\chi_1}  |\chi_1|^4  + \lambda_{\chi_2}  |\chi_2|^4+ \lambda_{\chi_3}  |\chi_3|^4+ \lambda_{\chi_4}  |\chi_4|^4+ \lambda_{\chi_5}  |\chi_5|^4+ \lambda_{\chi_6}  |\chi_6|^4+ \lambda_{\chi_7}  |\chi_7|^4   \\ \nonumber 
&&+     \lambda_{\varphi \chi_{ij}}  \varphi^\dagger \varphi   \chi^\dagger_i \chi_j  + \lambda_{\chi_{12}}  |\chi_1|^2 |\chi_2|^2  + \lambda_{\chi_{13}}  |\chi_1|^2 |\chi_3|^2  + \lambda_{\chi_{14}}  |\chi_1|^2 |\chi_4|^2  + \lambda_{\chi_{15}}  |\chi_1|^2 |\chi_5|^2   \\ \nonumber
&&+ \lambda_{\chi_{16}}  |\chi_1|^2 |\chi_6|^2  + \lambda_{\chi_{17}}  |\chi_1|^2 |\chi_7|^2 
+ \lambda_{\chi_{23}}  |\chi_2|^2 |\chi_3|^2  + \lambda_{\chi_{24}}  |\chi_2|^2 |\chi_4|^2  + \lambda_{\chi_{25}}  |\chi_2|^2 |\chi_5|^2  + \lambda_{\chi_{26}}  |\chi_2|^2 |\chi_6|^2   \\ \nonumber
&&+ \lambda_{\chi_{27}}  |\chi_2|^2 |\chi_7|^2  
+ \lambda_{\chi_{23}}  |\chi_2|^3 |\chi_3|^4  + \lambda_{\chi_{25}}  |\chi_2|^3 |\chi_5|^2  + \lambda_{\chi_{36}}  |\chi_3|^2 |\chi_6|^2  + \lambda_{\chi_{37}}  |\chi_3|^2 |\chi_7|^2   \\ \nonumber
&&+ \lambda_{\chi_{45}}  |\chi_4|^2 |\chi_5|^2  + \lambda_{\chi_{46}}  |\chi_4|^2 |\chi_6|^2  + \lambda_{\chi_{47}}  |\chi_4|^2 |\chi_7|^2 + \lambda_{\chi_{56}}  |\chi_5|^2 |\chi_6|^2  + \lambda_{\chi_{57}}  |\chi_5|^2 |\chi_7|^2  \\ \nonumber
&&  + \lambda_{\chi_{67}}  |\chi_6|^2 |\chi_7|^2  + \lambda_{1122}  \chi_1^2 \chi_2^2  
+ \lambda_{2224}  \chi_2^3 \chi_4   + \lambda_{2555}  \chi_2 \chi_5^{\dagger 3} 
+ \rm H.c..
\label{SP} 
\eea

We can parametrize the scalar fields as,
\begin{align}
 \chi_r(x) &=\frac{v_r + s_r(x) +i\, a_r(x)}{\sqrt{2}}, ~ \varphi =\left( \begin{array}{c}
G^+ \\
\frac{v+h+i G^0}{\sqrt{2}} \\
\end{array} \right).
\end{align}
Moreover, we assume that all quartic couplings are simply order 1 in the scalar potential.  Furthermore, for simplifying the scalar potential and phenomenology, we set $\lambda_{1122} = \lambda_{2224} =\lambda_{2555}=0$.  This choice  is insensitive to the phenomenological results presented in this work.

For studying  potential collider signatures of the SHVM at hadron colliders,  we  introduce a softly broken scalar potential\footnote{The soft-symmetry breaking parameters $\rho$ and $\sigma$ are assumed to be independent of the scale $\Lambda$.  This is possible in the UV completion of the SHVM discussed in the appendix, where the parameters $\rho$ and $\sigma$ originate from the DTC dynamics, and the scale $\Lambda$ corresponds to the scale of the dark-QCD. For more details, see Ref. \cite{Abbas:2025ser}. } given by,
\begin{align}
V_{\rm soft}
= -\rho_r^2 \ \chi_r^2  + \sigma_r  \ \chi_r^3   \rm +  H.c.,
\label{soft_pot}
\end{align}
where the soft-breaking terms having $\rho_r$ and $\sigma_r$ parameters are added for all scalars $(r=1-7)$.

A global symmetry is softly broken when the dimension-four part of the Lagrangian respects the symmetry, but lower-dimensional operators of dimension two or three do not. Such quadratic or cubic terms, with coefficients of mass dimension two or one, preserve renormalizability~\cite{Symanzik:1969ek,Symanzik:1970zz}, since no new counter-terms beyond those of the symmetric theory are required. The soft-breaking term permits the extra scalar masses to be arbitrarily large, allowing the model to remain consistent with the absence of new physics signals at the LHC~\cite{Ferreira:2022gjh}.

Using $v_r = \sqrt{2} \epsilon_r \Lambda$, we can write the masses of scalars  approximately as,
\begin{align}
\label{scal_mass}
m_{s_r}^2 \approx  & 2 \epsilon_r \Lambda \left(8 \epsilon_r \Lambda  +3 \sqrt{2} \sigma_r \right).
\end{align}

The pseudoscalar mass-matrix is completely diagonal, and the masses of pseudoscalars are given by
\begin{align}
\label{ps_mass}
m_{a_r}^2 = & 4 \rho_r^2 -  9 \sqrt{2} \epsilon_r \Lambda \sigma_r.
\end{align}

The condition that $m_{a_r}^2  \geq 0$, implies that,
\begin{align}
 4 \rho_r^2 \geq 9 \sqrt{2} \epsilon_r \Lambda \sigma_r.
\end{align}

We investigate a scenario where the masses of scalars and pseudoscalars are hierarchical and almost degenerate, and mixing among them is effectively negligible.   We notice that the soft-symmetry breaking parameters $\sigma_r$ are mass parameters.  Therefore, it would be fair to assume that,
\begin{align}
    \sigma_r \propto v_r,
\end{align}
leading to
\begin{align}
\label{sig_r}
    \sigma_r = k_r  v_r = k_r \sqrt{2} \epsilon_r \Lambda,
\end{align}
where $k_r$ is the proportionality constant and $r=1-7$.

The diagonalization of the squared scalar mass matrix is performed  as
\begin{align}
    U^T \mathcal{M}_s^2 U= \mathcal{M}_{dia}^2.  
\end{align}
Using equations \ref{epsi},  \ref{sig_r},  $k_{1-6}= 10^7$, we can compute the rotation matrix $U$ as,
\begin{equation}
U = \begin{pmatrix}
1 & 0 & 0 & 0 & 0 & 0 & 0 \\
0 & -1 & -3.46\times10^{-9} & -7.81\times10^{-10} & 0 & 0 & 0 \\
0 & 3.46\times10^{-9} & -1 & -1.59\times10^{-8} &
-1.34\times10^{-9} & -7.53\times10^{-10} & 0 \\
0 & 7.81\times10^{-10} & 1.59\times10^{-8} & -1 &
-5.95\times10^{-9} & -3.34\times10^{-9} & 0 \\
0 & 0 & 1.34\times10^{-9} & 5.95\times10^{-9} & -1 &
-5.52\times10^{-8} & 2.34\times10^{-10} \\
0 & 0 & 7.53\times10^{-10} & 3.34\times10^{-9} &
5.52\times10^{-8} & -1 & 4.16\times10^{-10} \\
0 & 0 & 0 & 0 & 2.34\times10^{-10} & 4.16\times10^{-10} & 1
\end{pmatrix}
\label{U-mat}
\end{equation}
Thus, we see that the rotation matrix $U$ is practically a unit matrix.  Therefore, the scalars $s_i$ effectively do not mix.   The above result does not depend on the value of $k_7$.

The tree-level contributions of the scalars and pseudoscalars to the neutral meson mixing are given by the  Wilson coefficients  \cite{Buras:2013rqa,Crivellin:2013wna},
\begin{align}
C_2^{ij} &= -(y_{ji}^*)^2\left(\frac{1}{m_{s_i}^2}-\frac{1}{m_{a_i}^2}\right),\notag \\
\tilde C_2^{ij} &= -y_{ij}^2\left(\frac{1}{m_{s_i}^2}-\frac{1}{m_{a_i}^2}\right),\notag \\
C_4^{ij} &= -\frac{y_{ij}y_{ji}}{2}\left(\frac{1}{m_{s_i}^2}+\frac{1}{m_{a_i}^2}\right),
\label{eq:wilsons}
\end{align}
where $m_{s_i}$ and $m_{a_i}$ are the masses of scalar and pseudoscalar particles of the SHVM, respectively.

From the down-type mass matrix given in equation \ref{mUD}, we notice that the SHVM  contribution to $B_{d,s}$-mixing and $B_{d,s} \rightarrow \mu^+ \mu^-$ decays do not exist.  However, the neutral kaon mixing and   $K_L \rightarrow \mu^+ \mu^-$ decay still get contribution from the SHVM.  We notice that for a almost degenerate choice  $m_{s_4} \approx m_{a_4}$,  the contribution to the Wilson coefficients $C_2^{ij}$ and $\tilde C_2^{ij}$ of the neutral kaon mixing from the SHVM, vanishes.  The operator  $\bar{s}_R d_L \bar{s}_L d_R $ corresponding to the  Wilson coefficient $C_4^{ij}$, does not exist.  Thus, the SHVM is free from any constraint coming from the neutral kaon mixing.

For   $m_{s_4} \approx m_{a_4}$,  the SHVM does not contribute to $K_L \rightarrow \mu^+ \mu^-$ decay.  This can be understood by noting that the scalars  in the SHVM effectively do not mix.  This decay can arise through the operator $\bar{d}_L s_R$, which couples to the singlet field $\chi_4$.  However, the operator $\mu_{L}\mu_{R}$ couples to the singlet field $\chi_5$.  Since the mixing among scalars  are effectively negligible, this decay does not impose any constraint on the parameter space of the SHVM.

\subsection{Bounds from lepton flavour violation}
We now investigate if the leptonic flavor violation can place bounds on the SHVM.  The most important bounds could come from radiative leptonic decay $\mu \rightarrow  e\gamma$, which appears at  loop-level.  However,  this process does not receive any contribution from the SHVM due to an effective absence of mixing among the scalars.  

The physical scalar mass eigenstates $h_i$ are related to the unphysical scalar sates $s_i$ through the matrix in Eq. \ref{U-mat}, and given by,
\begin{align}
\mathrm{matrix}(s_i)^T = U ~\mathrm{matrix}(h_i)^T.
\end{align}
We notice that the matrix $U $ is effectively diagonal due to negligible off-diagonal elements for $k_{1-6}= 10^7$.  Therefore, $s_i = h_i$.  The mass matrix of  pseudoscalars are diagonal.  Therefore, they do not contribute to lepton flavor violating processes.  

The effective Lagrangian for the radiative leptonic decays can be written as,
\begin{align}
\lag_{\text{eff}}&=m_{\ell'}\, C_T^L\,\bar \ell \sigma^{\rho\lambda}P_L\,\ell' \,F_{\rho\lambda}+m_{\ell'}\, C_T^R\,\bar \ell \sigma^{\rho\lambda}P_R\,\ell'\,F_{\rho\lambda}.
\label{meg}
\end{align}

The branching ratio of the radiative leptonic decays turns out to be,
\begin{align}
\br(\ell'\rightarrow  \ell\gamma)=\frac{m_{\ell'}^5}{4\pi \Gamma_{\ell'}}\left(|C_T^L|^2+|C_T^R|^2\right).
\end{align}

The one-loop contribution can be parametrized in terms of the following Wilson coefficients  \cite{Bauer:2015kzy},
\begin{align}
C_T^L = (C_T^R)^*
=\frac{e}{32\pi^2}\sum_{k=e,\mu,\tau} &\bigg\{ \frac{1}{6}\left( \,y^*_{\ell k}y_{\ell' k}+\frac{m_\ell}{m_k}y^*_{k \ell }y_{k\ell' }\right)\left(\frac{1}{m_s^2}-
\frac{1}{m_a^2}\right)\notag\\
&-y_{\ell k}y_{k\ell'}\frac{m_k}{m_{\ell'}}
\left[ \frac{1}{m_s^2}
\left(\frac{3}{2}+\log\frac{m_{\ell'}^2}{m_s^2}\right)-\frac{1}{m_a^2}
\left(\frac{3}{2}+
\log\frac{m_{\ell'}^2}{m_a^2}
\right)\right]
\bigg\}.
\label{megw}
\end{align}

The current and future projected sensitivities of radiative leptonic decay $\mu \rightarrow  e\gamma$ are taken from the reference \cite{Abbas:2024dfh}, and are given by,
\begin{align}
BR(\mu \rightarrow  e\gamma)  < 4.2 \times 10^{-13}  \hspace{0.16cm}  \text{[MEG],}  \nonumber \\
BR(\mu \rightarrow  e\gamma)  < 6 \times 10^{-14}  \hspace{0.16cm}  \text{[MEG -II].}
\label{meg}
\end{align}

\begin{figure}[H]
 	\centering
  \begin{subfigure}[]{0.46\linewidth}
     \includegraphics[width=\linewidth]{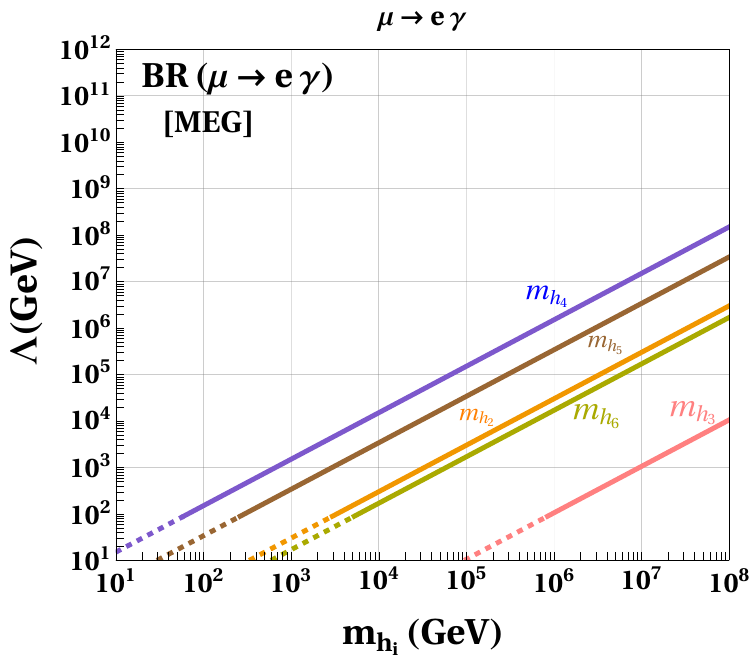}
     \caption{}
          \label{figmeya}	
 \end{subfigure}
 \begin{subfigure}[]{0.46\linewidth}
     \includegraphics[width=\linewidth]{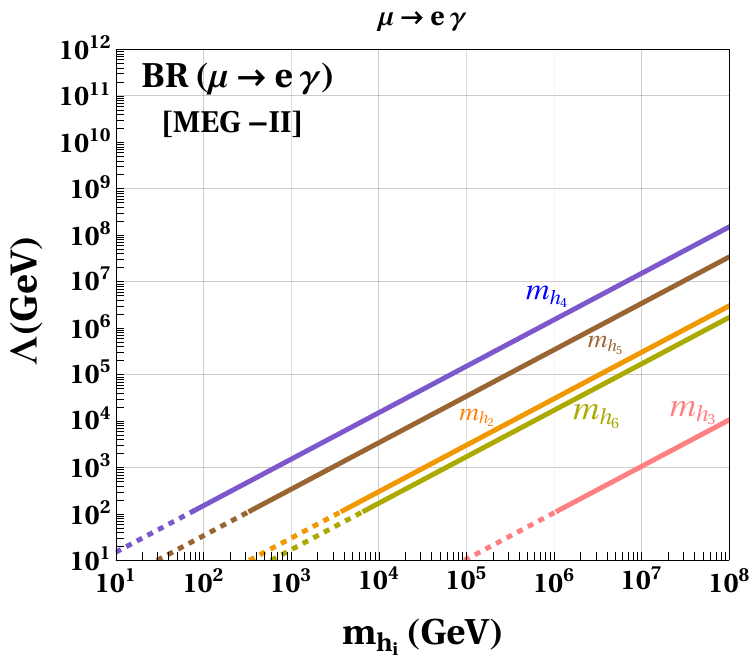}
     \caption{}
         \label{figmeyb}	
\end{subfigure}
 \caption{Constraints on the masses of the scalars $m_{h_i}$ and the scale $\Lambda$ from the experimental upper limits of radiative leptonic decay $\mu \rightarrow  e\gamma$, as given in the equation \ref{meg}. The solid lines represent the allowed regions, while the dashed lines indicate the excluded parameter space.     }
  \label{figmey}
	\end{figure}

The bounds on the masses of the scalars $m_{h_i}$ and the scale $\Lambda$ from the branching ratio of the radiative leptonic decays are  shown in figure \ref{figmey} for  $k_{1-6}= 10^7$. We notice the current sensitivity from MEG and  the projected limits of the MEG-\rom{2} experiments do not impose any constraints on the scale $\Lambda$ beyond 100 GeV for the masses $m_{h_{2,3,4,5,6}}$.  The mass $m_{h_1}$ is insensitive to $\mu \rightarrow  e\gamma$ limits.

\subsection{Couplings of scalars and pseudoscalars $a_r$ to fermions}
We can write the singlet  fields $\chi_r$ as,
\begin{equation}
\label{eps}
\frac{\chi_r}{\Lambda} = \epsilon_r  [ 1 +  \frac{s_r + i a_r}{v_r} ].
\end{equation}

The  equation \ref{mass2} can now be written  in the following form:
\begin{equation}
 y_{ij}^f  \frac{\chi_r}{\Lambda}   \varphi  \bar{f}f
\cong  y_{ij}^f  \epsilon_r  \frac{v}{\sqrt{2}} \left[1 + \frac{ (s_r +  i  a_r )}{v_r} + \frac{h}{v}\right] \bar{f} f,
\label{couplings}
\end{equation}
where $f= u,d,\ell$.

The couplings of  Goldstone bosons $a_r$  with fermions $\bar{f} f$ are now given by,
\begin{eqnarray}
\label{coupl_mat}
y_{a_r f_{iL} f_{jR}}^{u} \equiv  y_{a_r ij}^u & =& i  \frac{v}{\sqrt{2}} 
\begin{pmatrix}
 y_{11}^u  \epsilon_1/v_1 &  0  &   y_{13}^u \epsilon_{5}/v_5    \\
0    &  y_{22}^u \epsilon_{2}/v_2  &   y_{23}^u \epsilon_{2}/v_2  \\
 0  &  0  &   y_{33}^u \epsilon_{3}/v_3
\end{pmatrix}, \\ \nonumber
y_{a_r ij}^{d} & = & i \frac{v}{\sqrt{2}} 
\begin{pmatrix}
 y_{11}^d  \epsilon_{1}/v_1 &  y_{12}^d \epsilon_{4}/v_4  &   0    \\
 0    &  y_{22}^d \epsilon_{5}/v_5  &   0  \\
 0  &  0  &   y_{33}^d \epsilon_{6}/v_6
\end{pmatrix},    \\ \nonumber
y_{a_r ij}^{\ell} &=& i \frac{v}{\sqrt{2}} 
\begin{pmatrix}
 y_{11}^\ell  \epsilon_1/v_1 &  y_{12}^\ell \epsilon_{4}/v_4  &   y_{13}^\ell \epsilon_{5}/v_5    \\
 0    &  y_{22}^\ell \epsilon_{5}/v_5  &   y_{23}^\ell \epsilon_{2}/v_2  \\
 0  &  0  &   y_{33}^\ell  \epsilon_{2}/v_2
\end{pmatrix}.
\end{eqnarray}

We notice that due to the appearance of the factor $1/v_r$ in the coupling matrices, the FCNC interactions of scalars and  pseudoscalars  with fermions are unavoidable. The coupling of scalars to fermions are determined by using equation \ref{couplings}, and bi-unitary transformation matrices used for diagonalizing the fermionic mass matrices.

Now using  $v_r = \sqrt{2} \epsilon_r \Lambda$, the coupling matrices in equation \ref{coupl_mat}, can be written in a simpler form,

\begin{eqnarray}
\label{coupl_mat2}
y_{a_r f_{iL} f_{jR}}^{u} \equiv  y_{a_r ij}^u & =& i  \frac{v}{2 \Lambda} 
\begin{pmatrix}
 y_{11}^u   &  0  &   y_{13}^u    \\
0    &  y_{22}^u   &   y_{23}^u   \\
 0  &  0   &   y_{33}^u 
\end{pmatrix}, \\ \nonumber
y_{a_r ij}^{d} & = & i  \frac{v}{2 \Lambda} 
\begin{pmatrix}
 y_{11}^d  &  y_{12}^d   &   0    \\
 0    &  y_{22}^d   &   0  \\
0   & 0   &   y_{33}^d 
\end{pmatrix},    \\ \nonumber
y_{a_r ij}^{\ell} &=& i  \frac{v}{2 \Lambda}
\begin{pmatrix}
 y_{11}^\ell   &  y_{12}^\ell   &   y_{13}^\ell    \\
 0    &  y_{22}^\ell   &   y_{23}^\ell   \\
 0  &  0  &   y_{33}^\ell 
\end{pmatrix}.
\end{eqnarray}

\section{Collider physics of the SHVM}
 \label{collider}
In this section, we shall discuss the inclusive collider signatures of the SHVM. For computing  the production cross-sections of scalars $s_i/a_i$  in various channels, we use the {\tt MSTW2008} PDF \cite{Martin:2009iq}.  We investigate inclusive signatures $ pp \rightarrow s_r/a_r \rightarrow f_i f_j/\gamma \gamma$ of the SHVM at the HL-LHC, HE-LHC, and a 100 TeV collider.  We observe that due to the absence of interactions between the Higgs and $\chi_r$ fields in this work, there are no bounds on the masses of scalars and pseudo-scalars from the direct LHC searches \cite{CMS:2018amk}.  Moreover, the decays of scalars and pseudoscalars to WW and ZZ  bosons  arise at  one-loop level.  Therefore, they are highly suppressed. ATLAS has placed a limit on masses above 300 GeV \cite{ATLAS:2022eap}, and the CMS limit is 200 GeV \cite{CMS:2019bnu}.  In the di-photon channel searches,  the ATLAS has placed a bound above 200 GeV \cite{ATLAS:2021uiz}, and that of the CMS is above 500 GeV \cite{CMS:2018dqv}.

The production cross section for producing a scalar $\phi=s_r$  or a pseudoscalar $a_r$, of mass $M$, which decays to a final state  X, can be written as \cite{Arcadi:2023smv},
\begin{equation}
\sigma(p p \rightarrow \phi \rightarrow X) =  \frac{1}{M s} \sum_j C_j\, \Gamma(\phi \to j)\, \text{BR}(\phi \to X),
\end{equation}
where $j$ represents the possible initial states for production of scalar or a pseudoscalar, $C_{j}$ denote the corresponding weight factor, and $s$ is the squared center of mass energy. The weight factors  $C_{j}$ corresponding to different initial states are given by \cite{Arcadi:2023smv},
\begin{equation}
C_{gg} = \frac{\pi^2}{8} \int_{M^2/s}^1 \frac{dx}{x} g(x)g\left(\frac{M^2}{sx}\right),
\end{equation}

\begin{equation}
C_{q\bar{q}} = \frac{4\pi^2}{9} \int_{M^2/s}^1 \frac{dx}{x} \left( q(x)\, \bar{q}\left( \frac{M^2}{sx} \right) + \bar{q}(x)\, q\left( \frac{M^2}{sx} \right) \right).
\end{equation}

The partial decay widths of the scalar and  pseudoscalar to $f\bar{f}$  can be written as \cite{Arcadi:2023smv},
\begin{eqnarray}
\Gamma(s_r\rightarrow ff) &=& \frac{N_C^f g_{sff}^2 m_f^2(M) M}{8\pi v^2}\left(1-\frac{4m_f^2}{M^2}\right)^{3/2},\\
\Gamma(a_r\rightarrow ff) &=& \frac{N_C^f g_{pff}^2 m_f^2(M) M}{8\pi v^2}\left(1-\frac{4m_f^2}{M^2}\right)^{1/2},
\end{eqnarray}
where $g_{sff},g_{pff}$ show the ratios of the quark coupling to the spin-0 particle and the SM Yukawa couplings,  and the color factor $N_C^f=3$ for quarks and 1 for leptons.

The partial decay widths to $gg$ and $\gamma\gamma$ are given by \cite{Arcadi:2023smv},
\begin{eqnarray}
\Gamma(s_r\rightarrow gg) &=& \frac{ \alpha_s^2 M^3}{32\pi^3v^2} |\sum_f g_{sff} F_S\left(\frac{M^2}{4m_f^2}\right)|^2,\\
\Gamma(a_r\rightarrow gg) &=& \frac{ \alpha_s^2 M^3}{32\pi^3 v^2} |\sum_f g_{pff} F_P\left(\frac{M^2}{4m_f^2}\right)|^2,\\
\Gamma(s_r\rightarrow \gamma\gamma) &=& \frac{ \alpha^2 M^3}{256\pi^3v^2} |\sum_f 2N_C^f Q_f^2 g_{sff} F_S\left(\frac{M^2}{4m_f^2}\right)|^2,\\
\Gamma(a_r\rightarrow \gamma\gamma) &=& \frac{ \alpha^2 M^3}{256\pi^3 v^2} |\sum_f 2N_C^f Q_f^2 g_{pff} F_P\left(\frac{M^2}{4m_f^2}\right)|^2.
\end{eqnarray}
The form factors $F_S(x)$ and $F_P(x)$ read,
\begin{eqnarray}
F_S(x) &=& x^{-1} (1+(1-x^{-1})f(x)),\\
F_P(x) &=&x^{-1}f(x).
\end{eqnarray}
where,
\[
f(x) =
\begin{cases}
\arcsin^2(\sqrt{x}), & x \leq 1 \\
-\dfrac{1}{4} \left( \log\left( \dfrac{\sqrt{x} + \sqrt{x - 1}}{\sqrt{x} - \sqrt{x - 1}} \right) - i\pi \right)^2, & x > 1
\end{cases}
\]
and $x= \frac{M^2}{4m_f^2}$.

\begin{table}[H]
\setlength{\tabcolsep}{6pt} 
\renewcommand{\arraystretch}{1} 
\centering
\begin{tabular}{l|cc|cc|cc}
\toprule
& \multicolumn{2}{c|}{$\mathcal{L}[fb^{-1}]$ [References]} & \multicolumn{2}{c|}{ATLAS 13 TeV} & \multicolumn{2}{c}{CMS 13 TeV} \\
$m_{a}$~[GeV] &  \myalign{c}{ATLAS} & \myalign{c|}{CMS} & \myalign{c}{500} & \myalign{c|}{1000} & \myalign{c}{500} & \myalign{c}{1000}   \\
\midrule
jet-jet ~[pb]          & $139$ \cite{ATLAS:2019fgd} &  $137$ \cite{CMS:2019gwf}  &  &0.1   &            &  0.2                 \\
$\tau \tau$~[pb]  & $36.1$ \cite{ATLAS:2017eiz} & 
 $35.9$ \cite{CMS:2018rmh}  & $8\e{-2}$ & $10^{-2}$  & $6\e{-2}$ & $10^{-2}$   \\
$e e$, $\mu \mu$~[pb]   & $139$ \cite{ATLAS:2019erb}&  
 $140$ \cite{CMS:2021ctt}  & $8\e{-4}$ &\phantom{xx}  $2\e{-4}$  & $2\e{-3}$ &\phantom{xx}  $4\e{-4}$  \\
$\mu e$~[pb]   & $138$ \cite{ATLAS:2020tre}&  $139$ \cite{CMS:2022fsw}  &  &\phantom{xx}  $3\e{-4}$  & $4\e{-3}$ &\phantom{xx}  $3\e{-4}$   \\
$ \mu \tau$~[pb]   & $138$ \cite{ATLAS:2020tre}&  $139$ \cite{CMS:2022fsw}  &  &\phantom{xx}  $1\e{-3}$  & $7\e{-3}$ &\phantom{xx}  $1\e{-3}$  \\
$e \tau$~[pb]   & $138$ \cite{ATLAS:2020tre}&  $139$ \cite{CMS:2022fsw}  &  &\phantom{xx}  $1\e{-3}$  & $5\e{-3}$ &\phantom{xx}  $1\e{-3}$  \\
$b  \bar{b}$~[pb]   & $139$ \cite{ATLAS:2019fgd}&  $138$ \cite{CMS:2022eud}  &  &\phantom{xx} $1\e{-2}$   &  &\phantom{xx} $4\e{-2}$   \\
$\gamma \gamma$~[pb]& $139$ \cite{ATLAS:2021uiz}  & $35.9$ \cite{CMS:2018dqv}  & $5\e{-4}$ & $1\e{-4}$  & $4\e{-3}$ &   $8\e{-4}$        \\
$t  \bar{t}$~[pb]& $36.1$ \cite{ATLAS:2019npw,ATLAS:2020lks}  & $35.9$ \cite{CMS:2018rkg} & $2\e{2}$ & $2$  & $30 $ &     $0.4$       \\
\bottomrule
\end{tabular}
\caption{Current limits for production cross section times branching ratio ($\sigma \times BR$) at 13 TeV LHC by ATLAS and CMS for high mass (pseudo)scalars resonance searches in inclusive production channels.}
\label{tab:limits_LHC}
\end{table}

In table \ref{tab:limits_LHC}, we show the current sensitivities of various inclusive production channels of (pseudo)scalars at the 14 TeV LHC.  To estimate the projected sensitivities at the HL-LHC, HE-LHC, and  a future 100 TeV collider for production of a heavy and a light pseudoscalar through these inclusive channels, we employ simple square-root scaling of the luminosity of the LHC as \cite{Abbas:2024dfh},
\begin{equation}
 \mathcal{S} \simeq \frac{S}{\sqrt{B}} \simeq    \sqrt{\mathcal{L}} \frac{\sigma_s}{\sqrt{\sigma_B}},
\end{equation} 
where $S$ is the number of signal events, $B$ denotes the background events, $\sigma_s$ is the signal cross-section, and 
$\sigma_B$ stands for the background cross-section.  The required cross-section of a signal at a future collider (FC) is given by \cite{Abbas:2024dfh}, 
\begin{align}
  \sigma_s^{\rm FC}  = \sqrt{\dfrac{\mathcal{L}_{\rm LHC}}{\mathcal{L}_{\rm FC}}} \sqrt{\dfrac{\sigma_{ B}^{ FC}}{\sigma_{ B}^{ LHC}}} \sigma_{s}^{\rm LHC},
\end{align}
where FC= HL-LHC, HE-LHC, and  a 100 TeV collider.  The  background cross-sections is given by   $\sigma_{ B}^{ FC} = 2 \sigma_{ B}^{ LHC}$  as discussed in ref. \cite{Abbas:2024dfh}.

The estimated  reaches ($\sigma \times BR$) of the HL-LHC, HE-LHC and a 100 TeV hadron collider in different inclusive channels for a heavy and a light pseudoscalar are presented in tables \ref{tab:futurelimits11} and \ref{tab:futurelimits_lighta1}. The background cross sections employed in these estimates are the same as those used in the reference ~\cite{Abbas:2024dfh}. In general, inclusive decay modes of the SHVM that are accessible to the HL-LHC,  the HE-LHC, and a 100~TeV collider,  are marked by the box (\fbox{}) in this work.

\begin{table}[H]
\setlength{\tabcolsep}{6pt} 
\renewcommand{\arraystretch}{1} 
\centering
\begin{tabular}{l|cc|cc|cc}
\toprule
 & \multicolumn{2}{c|}{HL-LHC [14 TeV, $3~\iab$] } & \multicolumn{2}{c|}{HE-LHC [27 TeV, $15~\iab$]} & \multicolumn{2}{c}{100 TeV, $30~\iab$} \\
$m_{a}$~[GeV] &  500 &  1000 &  500 &  1000 &  500 &  1000 \\
\midrule
$\tau \tau$ [pb]  & $7\e{-3}$ & $1\e{-3}$ & $4\e{-3}$ & $7\e{-4}$ & $5\e{-3}$ & $8\e{-4}$   \\
$e e$, $\mu \mu$ [pb] & $2\e{-4}$ & $4\e{-5}$  & $1\e{-4}$ & $3\e{-5}$  & $1\e{-4}$ & $3\e{-5}$  \\
$\mu e$ [pb]     &  $9\e{-4}$     & $7\e{-5}$ & $7\e{-4}$ & $5\e{-5}$  & $1\e{-3}$ & $1\e{-4}$  \\
$\mu \tau$ [pb]  &  $2\e{-3}$        &  $2\e{-4}$ & $1\e{-3}$ & $2\e{-4}$   &  $2\e{-3}$ & $3\e{-4}$  \\
$e \tau$ [pb]    & $1\e{-3}$           & $2\e{-4}$ & $8\e{-4}$ & $2\e{-4}$  &  $1\e{-3}$ & $3\e{-4}$  \\
$b  \bar{b}$ [pb]    & & $9\e{-3}$    &  & $5\e{-3}$    &  & $7\e{-3}$  \\
$\gamma \gamma$ [pb] & $1\e{-4}$  & $2\e{-5}$  & $6\e{-5}$ & $1\e{-5}$  & $7\e{-5}$ & $1\e{-5}$  \\
$t \bar{t}$ [pb]     & 4   & $5\e{-2}$      & 3      & $4\e{-2}$    & $8$ & $0.1$  \\
\bottomrule
\end{tabular}
\caption{Estimated reach ($\sigma \times BR$) of the HL-LHC, HE-LHC and a 100 TeV hadron collider for a high  mass pseudoscalar ($m_a$) in inclusive  production channels.}
\label{tab:futurelimits11}
\end{table}

We find that, for heavy scalar states, the present experimental sensitivities to lepton-flavour-violating modes are of the same order as those for the corresponding lepton-flavour-conserving channels involving $e$ and $\mu$ final states. A notable exception is the $\tau\tau$ and  $e (\mu) \tau$ final state, for which the sensitivities are  approximately two order and one order  of magnitude weaker, respectively.

 \begin{table}[H]
\setlength{\tabcolsep}{5pt} 
\renewcommand{\arraystretch}{0.7} 
\centering
\begin{tabular}{l|cc|cc|cc}
\toprule
 & \multicolumn{2}{c|}{HL-LHC [14 TeV, $3~\iab$] } & \multicolumn{2}{c|}{HE-LHC [27 TeV, $15~\iab$]} & \multicolumn{2}{c}{100 TeV, $30~\iab$} \\
$m_{a}$~[GeV] &  20 &  60 &  20 &  60 &  20 &  60 \\
\midrule
$\tau \tau$ [pb]  & & $0.9$ & & $0.5$ &  & $0.6$  \\
$\gamma \gamma$ [pb] & $1.3$  & $1.5$  & $0.7$ & $0.8$  & $0.8$ & $0.9$  \\
\bottomrule
\end{tabular}
\caption{Estimated reach ($\sigma \times BR$) of the HL-LHC, HE-LHC and a 100 TeV collider for low mass pseudoscalar ($m_a$) in inclusive  production channels.}
\label{tab:futurelimits_lighta1}
\end{table}

 \subsection{A unique signature of the SHVM }
A distinct signature of the SHVM, which keeps the SHVM apart from any other flavor models in the literature, is the decay of the pseudoscalar $a_3$.  The pseudoscalar $a_3$ couples to only $t\bar{t}$ pair.  Therefore, the only decay modes available for this particle are  $\gamma \gamma$, $gg$, and   $t\bar{t}$ pairs depending on the kinematics. Thus,  the pseudoscalar $a_3$ is a unique particle among pseudoscalars $a_i$.   

\begin{figure}[H]
	\centering
 \includegraphics[width=0.4\linewidth]{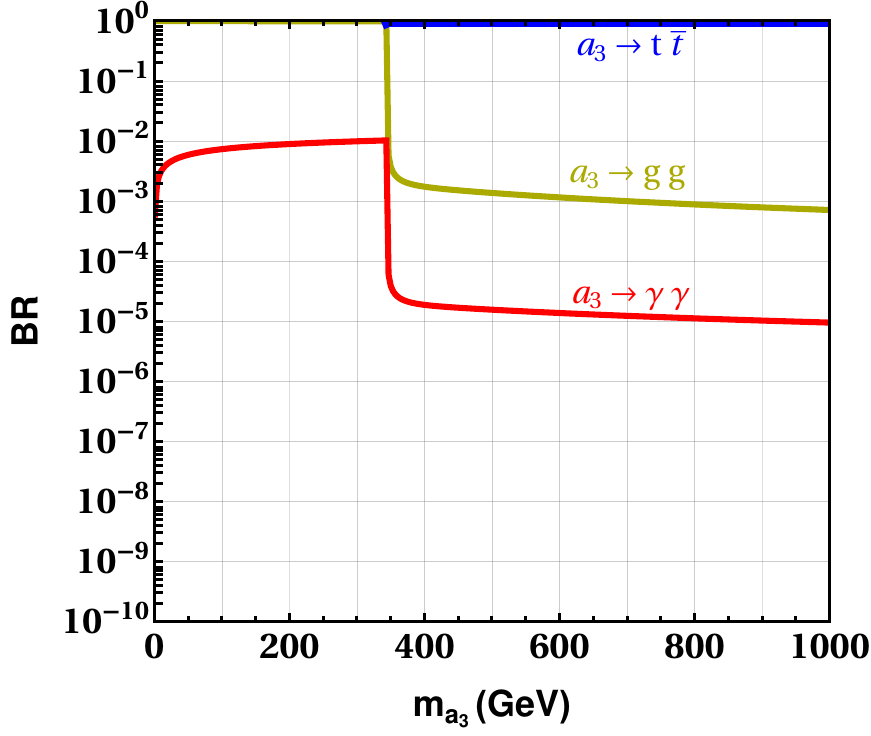}
 \caption{Branching ratios of various possible decay modes of the  pseudoscalar $a_3$. }
         \label{figa3br}
	\end{figure}

\begin{figure}[H]
	\centering
 \begin{subfigure}[]{0.327\linewidth}
 \includegraphics[width=\linewidth]{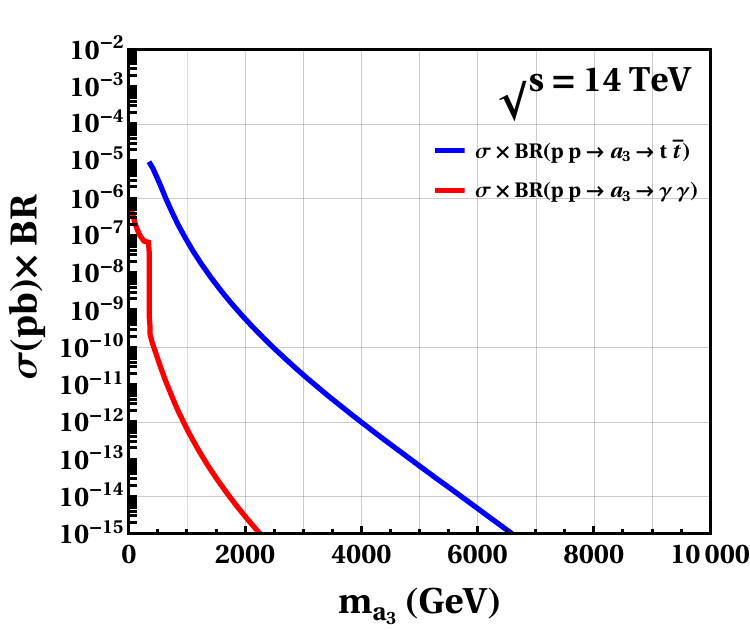}
 \caption{}
         \label{figa3a}
 \end{subfigure} 
 \begin{subfigure}[]{0.327\linewidth}
    \includegraphics[width=\linewidth]{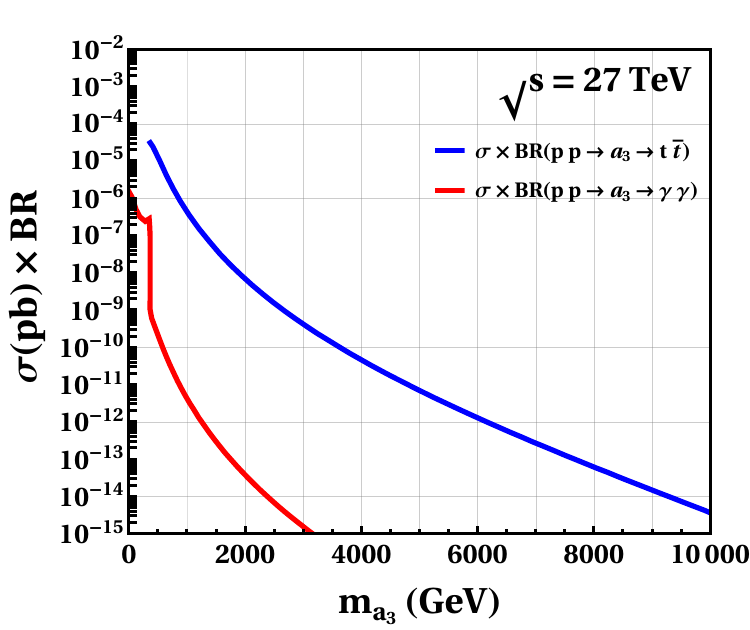}
    \caption{}
         \label{figa3b}	
\end{subfigure}
\begin{subfigure}[]{0.327\linewidth}
    \includegraphics[width=\linewidth]{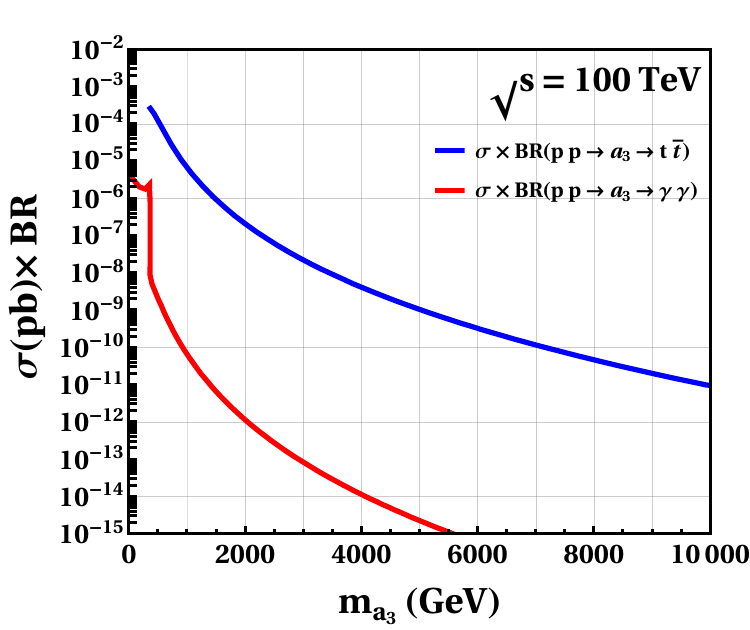}
    \caption{}
         \label{figa3c}	
\end{subfigure}
 \caption{Cross-section ($\sigma \times BR$) for the processes  $pp \rightarrow a_3 \rightarrow t \overline{t}$ and $pp \rightarrow a_3 \rightarrow \gamma \gamma$ as functions of the pseudoscalar mass $m_{a_3}$ are shown in figure \ref{figa3a} for the 14 TeV HL-LHC, figure \ref{figa3b} for the 27 TeV HE-LHC, and in figure \ref{figa3c} for a 100 TeV collider, where we have set the scale $\Lambda=500$ GeV.    }
  \label{figa3prod_a3}
	\end{figure}

In this section, we study the production and decays of pseudoscalar $a_3$ at the HL-LHC, HE-LHC, and a 100 TeV collider such as FCC-hh.  We show the branching ratios of the pseudoscalar $a_3$ in figure \ref{figa3br}, and the production cross sections for the $\gamma \gamma$ and $t\bar{t}$ modes, as a function of the mass of the pseudoscalar $a_3$,   are shown in figure \ref{figa3prod_a3} for the scale $\Lambda=500$ GeV.

We show in tables \ref{tab:futurelimita3a} and \ref{tab:futurelimita3b}, our benchmark predictions for the heavy and light pseudoscalar $a_3$ for the scale $\Lambda = 500$ GeV. We observe that the HL-LHC will not be able to search  heavy masses of pseudoscalar $a_3$.  The HE-LHC and a 100 TeV collider can probe the pseudoscalar $a_3$ in $t \bar{t}$  mode.  The light masses of pseudoscalar $a_3$ corresponding to the di-photon channel are beyond the reach of any colliders.

\begin{table}[H]
\setlength{\tabcolsep}{6pt} 
\renewcommand{\arraystretch}{1.2} 
\centering
\begin{tabular}{l|cc|cc|cc}
\toprule
 & \multicolumn{2}{c|}{HL-LHC [14 TeV, $3~\iab$] } & \multicolumn{2}{c|}{HE-LHC [27 TeV, $15~\iab$]} & \multicolumn{2}{c}{100 TeV, $30~\iab$} \\
$m_{a_3}$~[GeV] &  500 &  1000 &  500 &  1000 &  500 &  1000 \\
\midrule
$\gamma \gamma$ [pb] & $4\e{-6}$  & $6\e{-8}$  & $2\e{-5}$& $4\e{-7}$  & \fbox{$2\e{-4}$} & $6\e{-6}$  \\
$t \bar{t}$ [pb]     & $0.29$   & $7\e{-3}$      & $1.2$     & \fbox{$4\e{-2}$}    & \fbox{$12.6$} & $0.68$  \\
\bottomrule
\end{tabular}
\caption{Benchmark ($\sigma \times BR$) for higher masses of the pseudoscalar $m_{a_3}$ at $\Lambda = 500$ GeV, where $\rho_3 \simeq 7.1 \times 10^{4}$ GeV for $m_{a_3}= 500$ GeV and $m_{a_3}= 1000$ GeV, assuming $k_{1-6}=6 \times 10^3$.}
\label{tab:futurelimita3a}
\end{table}

\begin{table}[H]
\setlength{\tabcolsep}{5pt} 
\renewcommand{\arraystretch}{0.7} 
\centering
\begin{tabular}{l|cc|cc|cc}
\toprule
 & \multicolumn{2}{c|}{HL-LHC [14 TeV, $3~\iab$] } & \multicolumn{2}{c|}{HE-LHC [27 TeV, $15~\iab$]} & \multicolumn{2}{c}{100 TeV, $30~\iab$} \\
$m_{a_3}$~[GeV] &  20 &  60 &  20 &  60 &  20 &  60 \\
\midrule
$\gamma \gamma$ [pb] & $7\e{-2}$  & $4\e{-2}$  & $0.14$ & $8\e{-2}$  & $0.44$ & $0.37$  \\
\bottomrule
\end{tabular}
\caption{Benchmark ($\sigma \times BR$) for low  pseudoscalar mass $m_{a_3}$ at $\Lambda = 500$ GeV, where $\rho_3 \simeq 7.1 \times 10^{4}$ GeV for $m_{a_3}= 20$ GeV and $m_{a_3}= 60$ GeV, assuming $k_{1-6}=6 \times 10^3$.}
\label{tab:futurelimita3b}
\end{table}

\subsection{Inclusive production modes of scalars}
We now investigate the decay profiles and collider signatures emerging through the scalars $s_r$ of the SHVM, assuming the scale  $\Lambda = 500$ GeV.


\subsubsection{Scalar $s_1$}
\begin{figure}[H]
	\centering
    \includegraphics[width=0.45\linewidth]{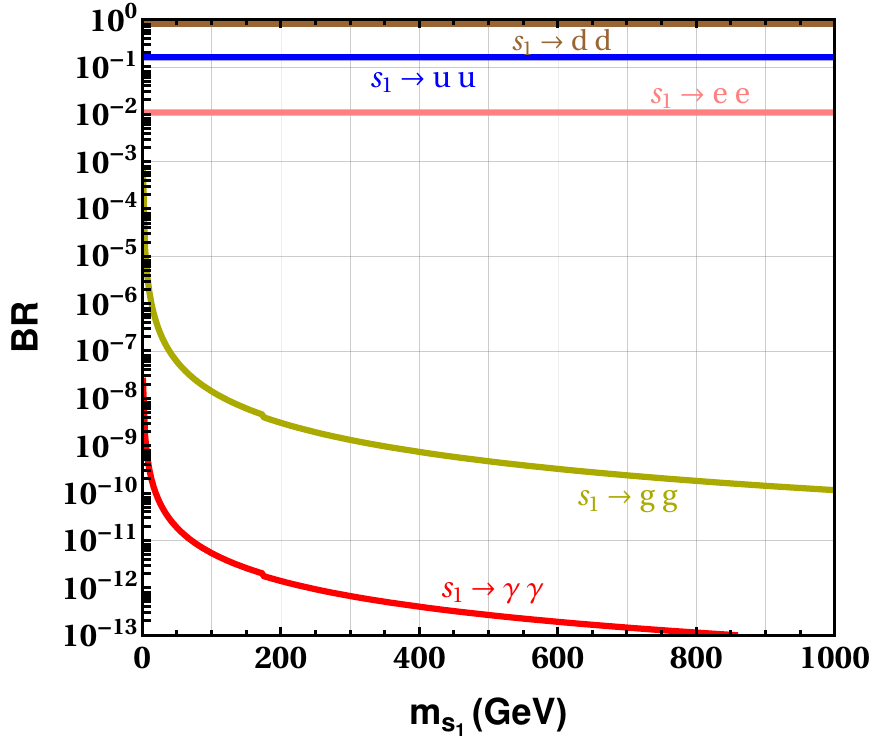}
 \caption{The branching ratio of various possible decay modes of the scalar $s_1$, assuming $\Lambda=500$ GeV.    }
  \label{figh1br}
	\end{figure}

\begin{figure}[H]
	\centering
 \begin{subfigure}[]{0.327\linewidth}
 \includegraphics[width=\linewidth]{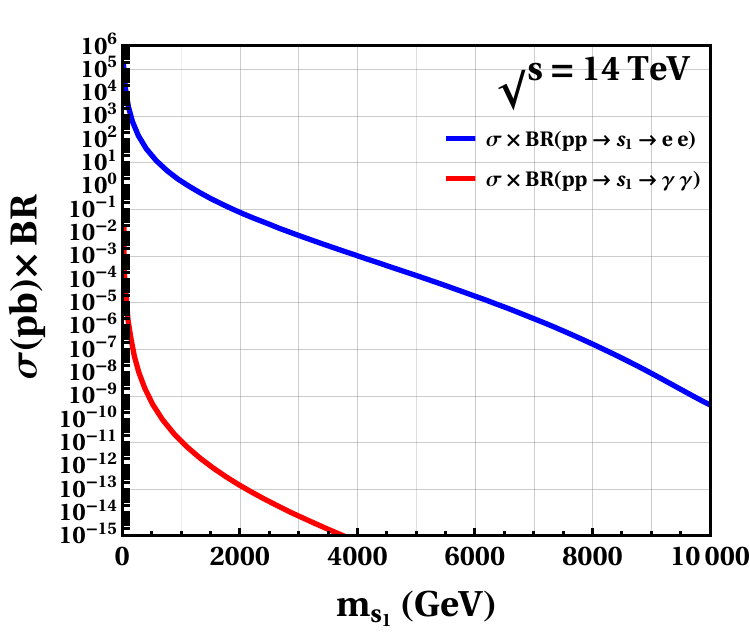}
 \caption{}
         \label{figh1a}
 \end{subfigure} 
 \begin{subfigure}[]{0.327\linewidth}
    \includegraphics[width=\linewidth]{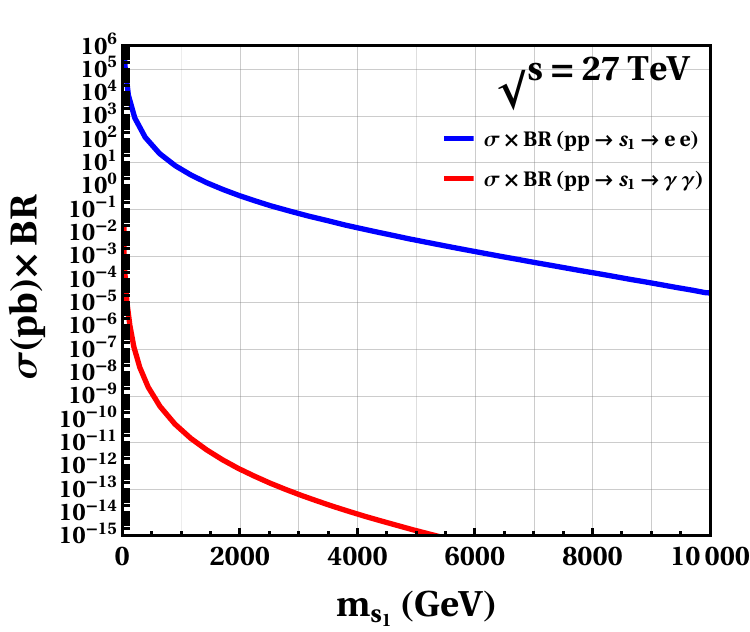}
    \caption{}
         \label{figh1b}	
\end{subfigure}
\begin{subfigure}[]{0.327\linewidth}
    \includegraphics[width=\linewidth]{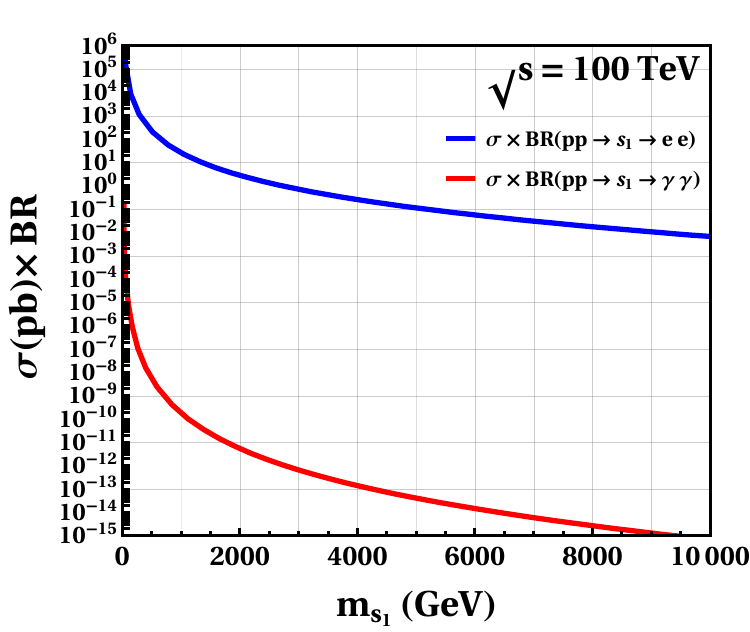}
    \caption{}
         \label{figh1c}	
\end{subfigure}
 \caption{ Cross-section ($\sigma \times BR$) for the processes  $pp \rightarrow s_1 \rightarrow ee$ and $pp \rightarrow s_1 \rightarrow \gamma \gamma$ as functions of the scalar mass $m_{s_1}$ are shown in figure \ref{figh1a} for the 14 TeV HL-LHC, figure \ref{figh1b} for the 27 TeV HE-LHC, and in figure \ref{figh1c} for a 100 TeV collider, where the scale $\Lambda=500$ GeV.    }
  \label{figh1sigma}
	\end{figure}

The dominant couplings of  scalar $s_1$ are with the pairs $e^- e^+$, $u\bar{u}$, and $d \bar{d}$.  We show the branching ratios of scalar $s_1$ in figure \ref{figh1br}.  The production cross-sections for inclusive modes $e^- e^+$ and $\gamma \gamma$ are shown in figure \ref{figh1sigma}.

The benchmark predictions for the scalar $s_1$ for heavy and light  masses at the scale $\Lambda = 500$ GeV are given in tables \ref{tab:futurelimith1a} and \ref{tab:futurelimith1b}. We notice that only $e^- e^+$ production channel is accessible to the HL-LHC, the HE-LHC, and a 100 TeV collider in the case of a heavy mass.

\begin{table}[H]
\setlength{\tabcolsep}{6pt} 
\renewcommand{\arraystretch}{1.2} 
\centering
\begin{tabular}{l|cc|cc|cc}
\toprule
 & \multicolumn{2}{c|}{HL-LHC [14 TeV, $3~\iab$] } & \multicolumn{2}{c|}{HE-LHC [27 TeV, $15~\iab$]} & \multicolumn{2}{c}{100 TeV, $30~\iab$} \\
$m_{s_1}$~[GeV] &  500 &  1000 &  500 &  1000 &  500 &  1000 \\
\midrule
$e \bar{e}$ [pb]     & \fbox{$21.2$}   & \fbox{$1.7$}      & \fbox{$49.7$ }    & \fbox{$4.9$}  & \fbox{$214.7$} & \fbox{$26.7$}  \\
$\gamma \gamma$ [pb] & $5\e{-10}$  & $1\e{-11}$  & $1\e{-9}$& $3\e{-11}$  & $5\e{-9}$ & $2\e{-10}$  \\
\bottomrule
\end{tabular}
\caption{Benchmark values of ($\sigma \times BR$) for  $m_{s_1}$ at $\Lambda = 500$ GeV. For $m_{s_1} = 500$ GeV, the proportionality constant $k_{1-6} = 8.3 \times 10^{9}$, and the corresponding value is $k_{1-6} = 3.3 \times 10^{10}$ for $m_{s_1} = 1000$ GeV.  }
\label{tab:futurelimith1a}
\end{table}

\begin{table}[H]
\setlength{\tabcolsep}{6pt} 
\renewcommand{\arraystretch}{1} 
\centering
\begin{tabular}{l|cc|cc|cc}
\toprule
 & \multicolumn{2}{c|}{HL-LHC [14 TeV, $3~\iab$] } & \multicolumn{2}{c|}{HE-LHC [27 TeV, $15~\iab$]} & \multicolumn{2}{c}{100 TeV, $30~\iab$} \\
$m_{s_1}$~[GeV] &  20 &  60 &  20 &  60 &  20 &  60 \\
\midrule
$\gamma \gamma$ [pb] & $2\e{-3}$  & $1\e{-5}$  & $3\e{-3}$& $3\e{-5}$  & $6\e{-3}$ & $8\e{-5}$  \\
\bottomrule
\end{tabular}
\caption{Benchmark ($\sigma \times BR$) for low values of the scalar mass $m_{s_1}$ at $\Lambda = 500$ GeV, where $k_{1-6}= 1.3 \times 10^{7}$ for $m_{s_1}= 20$ GeV, and $k_{1-6}= 1.2 \times 10^{8}$ for $m_{s_1}= 60$ GeV. }
\label{tab:futurelimith1b}
\end{table}

\subsubsection{Scalar $s_2$}
\begin{figure}[H]
	\centering
    \includegraphics[width=0.45\linewidth]{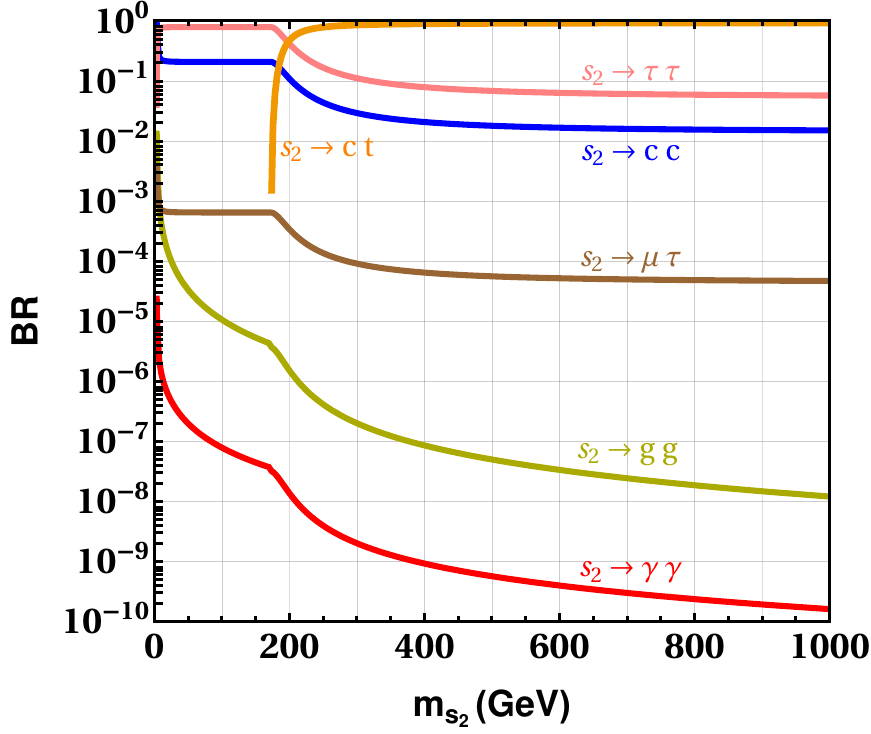}
 \caption{The branching ratio of various possible decay modes of the scalar $s_2$, assuming $\Lambda=500$ GeV.    }
  \label{figs2br}
	\end{figure}

\begin{figure}[H]
	\centering
 \begin{subfigure}[]{0.327\linewidth}
 \includegraphics[width=\linewidth]{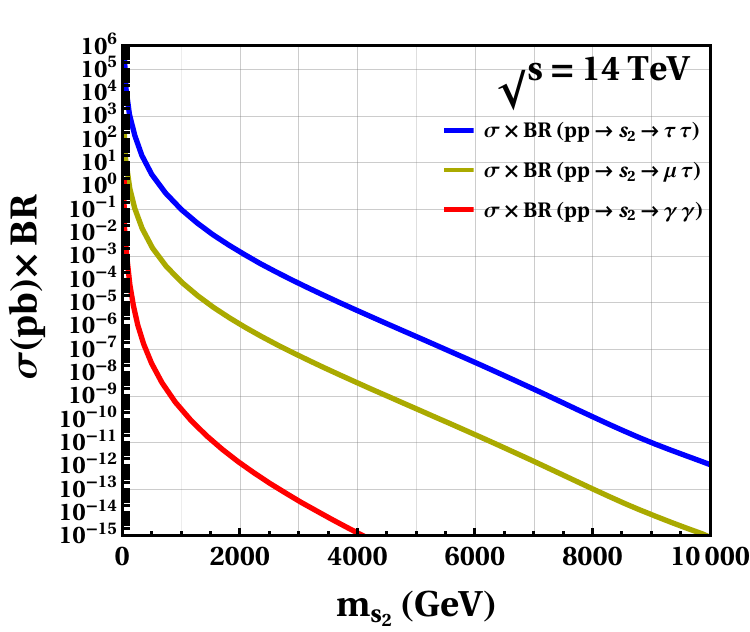}
 \caption{}
         \label{figh2a}
 \end{subfigure} 
 \begin{subfigure}[]{0.327\linewidth}
    \includegraphics[width=\linewidth]{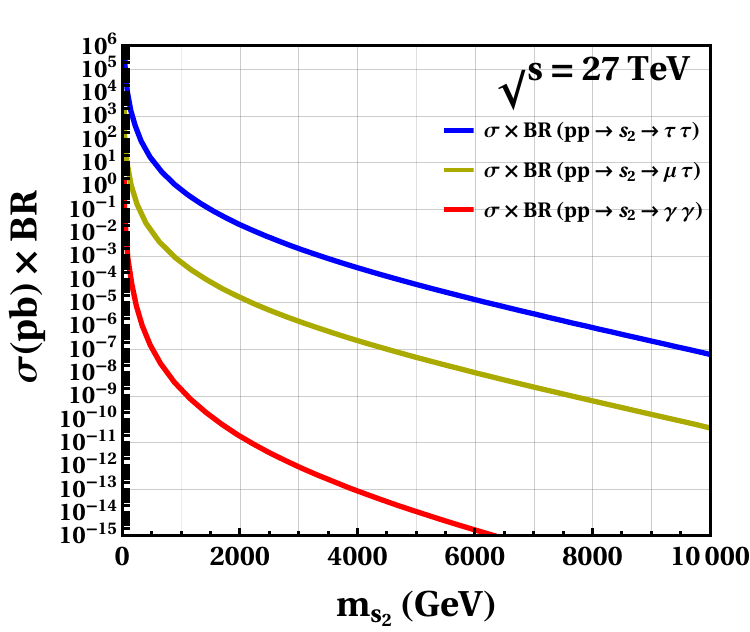}
    \caption{}
         \label{figh2b}	
\end{subfigure}
\begin{subfigure}[]{0.327\linewidth}
    \includegraphics[width=\linewidth]{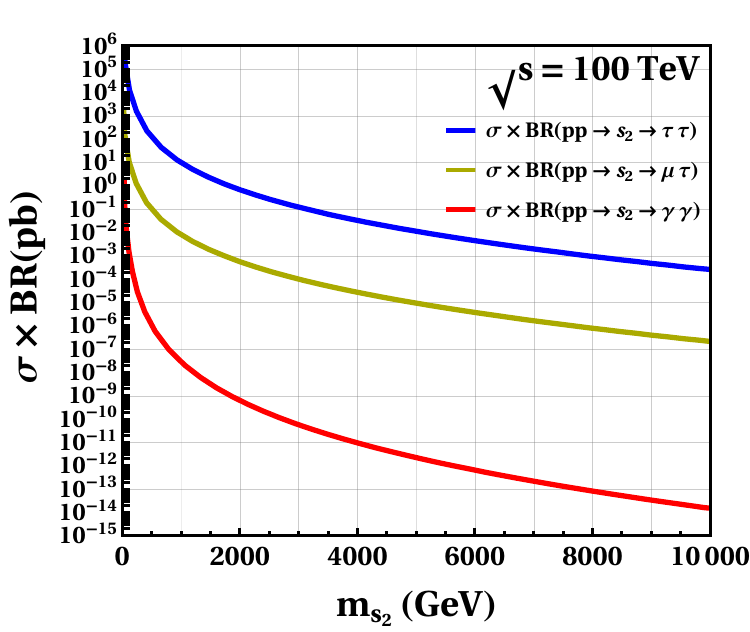}
    \caption{}
         \label{figh2c}	
\end{subfigure}
 \caption{Cross-section ($\sigma \times BR$) for the processes  $pp \rightarrow s_2 \rightarrow \tau \tau$, $pp \rightarrow s_2 \rightarrow \mu \tau$, and $pp \rightarrow s_2 \rightarrow \gamma \gamma$ as functions of the scalar mass $m_{s_2}$ are shown in figure \ref{figh2a} for the 14 TeV HL-LHC, \ref{figh2b} for the 27 TeV HE-LHC, and in figure \ref{figh2c} for a 100 TeV collider, where the scale $\Lambda=500$ GeV.       }
  \label{figh2sigma}
	\end{figure}

The scalar $s_2$ couples dominantly  with the pairs $\tau^- \tau^+$, $\mu \tau$,  $c\bar{c}$, and $c\bar{t}$.  We show the branching ratios of scalar $s_2$ in figure \ref{figs2br}.  The production cross-sections for inclusive modes $\tau^- \tau^+$,  $\mu^+ \tau^-$, and $\gamma \gamma$ at the HL-LHC, the HE-LHC, and a 100 TeV collider are shown in figure \ref{figh2sigma}.

The benchmark predictions for the scalar $s_2$ are shown in tables \ref{tab:futurelimith2a} for the heavy masses at the scale $\Lambda = 500$ GeV. The  $\tau \tau$ production channel is accessible to the HL-LHC, the HE-LHC and a 100 TeV collider.  The  $\mu^+ \tau^-$ mode is also within the reach of all colliders for a heavy mass.

 \begin{table}[H]
\setlength{\tabcolsep}{6pt} 
\renewcommand{\arraystretch}{1.2} 
\centering
\begin{tabular}{l|cc|cc|cc}
\toprule
 & \multicolumn{2}{c|}{HL-LHC [14 TeV, $3~\iab$] } & \multicolumn{2}{c|}{HE-LHC [27 TeV, $15~\iab$]} & \multicolumn{2}{c}{100 TeV, $30~\iab$} \\
$m_{s_2}$~[GeV] &  500 &  1000 &  500 &  1000 &  500 &  1000 \\
\midrule
$\tau \tau$ [pb]     & \fbox{$2.9$}   & \fbox{$0.1$}      & \fbox{$12.5$}     & \fbox{$0.7$}  & \fbox{$116$} & \fbox{$10.1$}  \\
$\mu \tau$ [pb]     & \fbox{$2.4\e{-3}$}   & $8.1\e{-5}$      & \fbox{$1\e{-2}$}     & \fbox{$5.4\e{-4}$}  & \fbox{$9.5\e{-2}$} & \fbox{$8.2\e{-3}$}  \\
$\gamma \gamma$ [pb] & $2.4\e{-8}$  & $2.7\e{-10}$  & $1\e{-7}$& $1.8\e{-9}$  & $9.6\e{-7}$ & $2.8\e{-8}$  \\
\bottomrule
\end{tabular}
\caption{Benchmark ($\sigma \times BR$) for higher values of the scalar mass $m_{s_2}$ at $\Lambda = 500$ GeV,  where $k_{1-6}= 8.9 \times 10^{3}$ for $m_{s_2}= 500$ GeV, and $k_{1-6}= 3.6 \times 10^{4}$ for $m_{s_2}= 1000$ GeV.}
\label{tab:futurelimith2a}
\end{table}

The scalar  $s_2$ can have a minimum mass of $4.09\times 10^2$ GeV, when the minimum value of the proportionality constant is $k_{1-6}=6 \times 10^3$.  Thus, the benchmark predictions for masses below $4.09\times 10^2$ GeV  of the scalar  $s_2$ are ruled out. 

\subsubsection{Scalar $s_3$}
The scalar $s_3$ primarily interacts with the $t \bar{t}$ pair at the tree level. For the minimum value of the  proportionality constant $k_{1-6}= 6 \times 10^3$, the mass of the scalar $s_3$ is approximately $116$ TeV at the scale   $\Lambda = 500$ GeV.  Thus, the scalar $s_3$ is not accessible at the HL-LHC, HE-LHC and even at a 100 TeV collider.

\subsubsection{Scalar $s_4$}
\begin{figure}[H]
	\centering
    \includegraphics[width=0.45\linewidth]{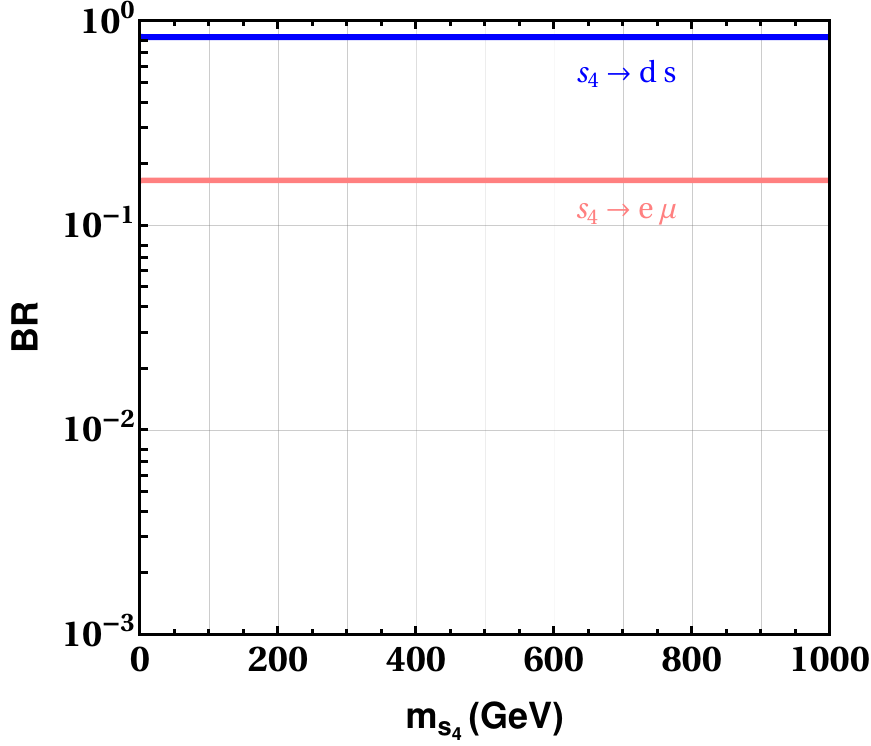}
 \caption{The branching ratio of various possible decay modes of the scalar $s_4$, assuming $\Lambda=500$ GeV.    }
  \label{figh4br}
	\end{figure}

\begin{figure}[H]
	\centering
 \begin{subfigure}[]{0.327\linewidth}
 \includegraphics[width=\linewidth]{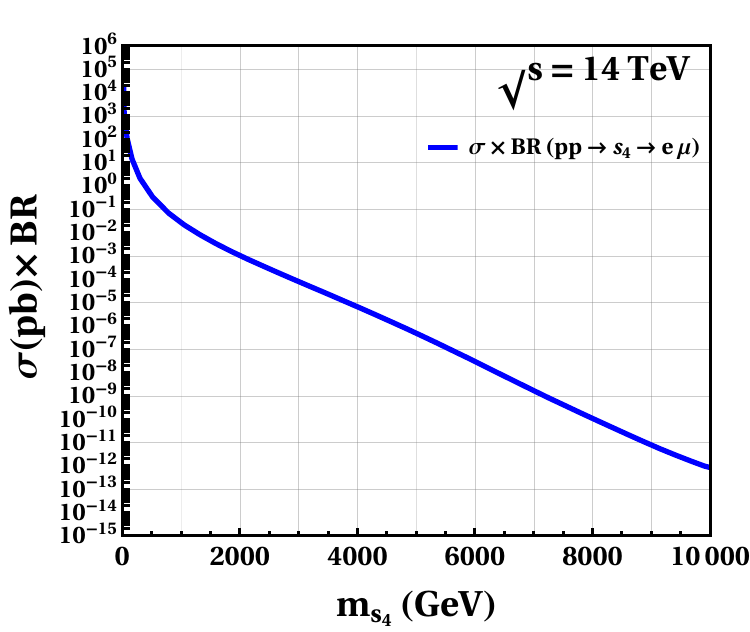}
 \caption{}
         \label{figh4a}
 \end{subfigure} 
 \begin{subfigure}[]{0.327\linewidth}
    \includegraphics[width=\linewidth]{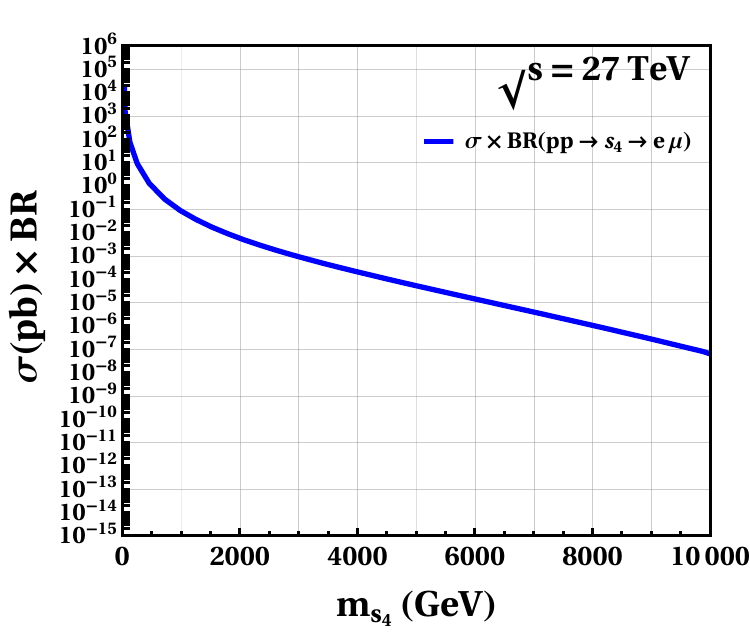}
    \caption{}
         \label{figh4b}	
\end{subfigure}
\begin{subfigure}[]{0.327\linewidth}
    \includegraphics[width=\linewidth]{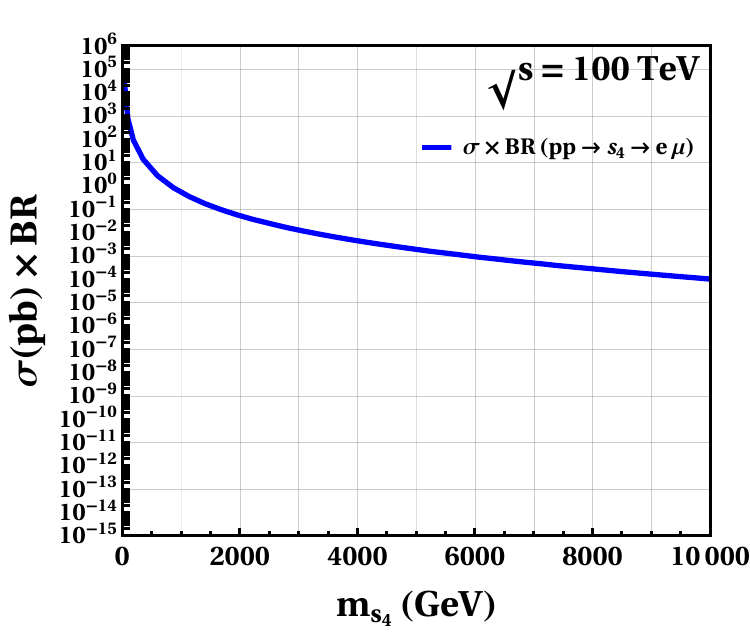}
    \caption{}
         \label{figh4c}	
\end{subfigure}
 \caption{Cross-section ($\sigma \times BR$) for the process  $pp \rightarrow s_4 \rightarrow e \mu$ as functions of the scalar mass $m_{s_4}$ are shown in figure \ref{figh4a} for the 14 TeV HL-LHC, figure \ref{figh4b} for the 27 TeV HE-LHC, and in figure \ref{figh4c} for a 100 TeV collider, where we have taken the scale $\Lambda=500$ GeV.    }
  \label{figh4sigma}
	\end{figure}

The possible decay channels of the scalar $s_4$ are represented in figure \ref{figh4br}, where it dominantly decays to the  $ \bar{d} s$ and $ \bar{e} \mu$ pairs. The scalar $s_4$ is produced through the annihilation of  $ \bar{d} s$, and its production cross-section for the inclusive channels $ \bar{e} \mu$ is  shown in figure \ref{figh4sigma} at the HL-LHC, HE-LHC, and a 100 TeV collider. Thus, the scalar $s_4$ represents another unique signature of the SHVM.

The benchmark predictions of the production cross-sections at the scale $\Lambda = 500$ GeV, through the inclusive channels $ \bar{e} \mu$  are given in table \ref{tab:benchh4_heavy} for the case when the scalar $s_4$ is heavy. It is evident that the inclusive channel $e \bar{\mu}$ is within the reach of HL-LHC, HE-LHC, and a 100 TeV collider for a heavy mass of scalar $m_{s_4}$.

\begin{table}[H]
\setlength{\tabcolsep}{6pt} 
\renewcommand{\arraystretch}{1.2} 
\centering
\begin{tabular}{l|cc|cc|cc}
\toprule
 & \multicolumn{2}{c|}{HL-LHC [14 TeV, $3~\iab$] } & \multicolumn{2}{c|}{HE-LHC [27 TeV, $15~\iab$]} & \multicolumn{2}{c}{100 TeV, $30~\iab$} \\
$m_{s_4}$~[GeV] &  500 &  1000 &  500 &  1000 &  500 &  1000 \\
\midrule
$e \mu$ [pb]     & \fbox{$0.37$}   & \fbox{$2.7\e{-2}$}      & \fbox{$0.93$}     & \fbox{$9\e{-2}$}  & \fbox{$4.51$} & \fbox{$0.53$}  \\
\bottomrule
\end{tabular}
\caption{Benchmark ($\sigma \times BR$) for higher values of the scalar mass $m_{s_4}$ at $\Lambda = 500$ GeV, where $k_{1-6}= 2.2 \times 10^{7}$ for $m_{s_4}= 500$ GeV, and $k_{1-6}= 8.9 \times 10^{7}$ for $m_{s_4}= 1000$ GeV.}
\label{tab:benchh4_heavy}
\end{table}

The lowest allowed value of the mass of $s_4$ is $8$ GeV for the minimum allowed value of the proportionality constant $k_{1-6}= 6 \times 10^{3}$.  However, for low mass values such as 20 and 60 GeV, the LHC does not have sensitivity for the signature of the $s_4$ in $\bar{e} \mu$ mode.

\subsubsection{Scalar $s_5$}
\begin{figure}[H]
	\centering
    \includegraphics[width=0.45\linewidth]{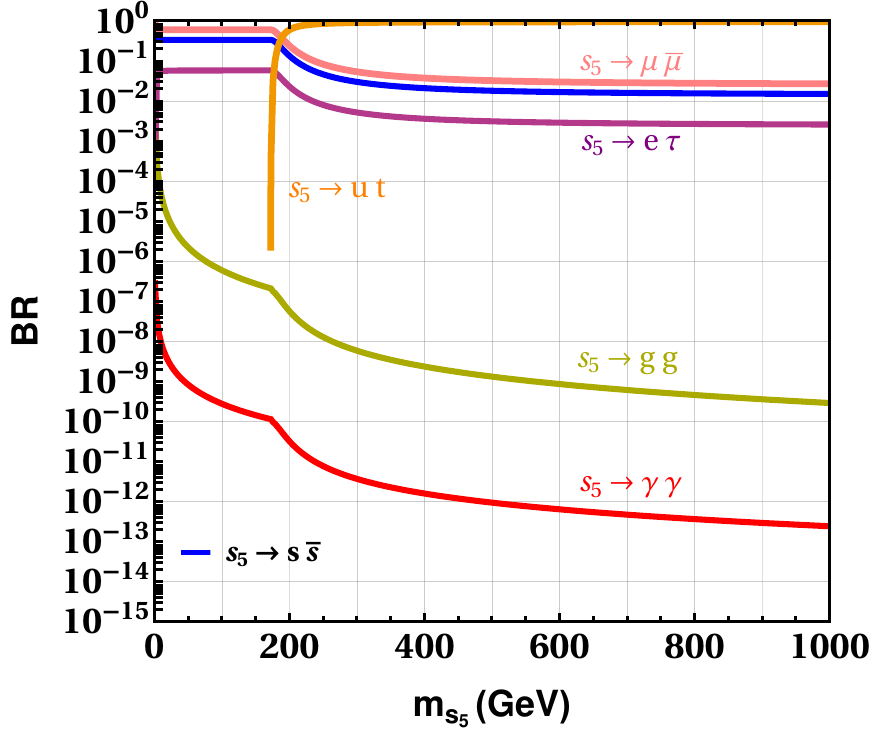}
 \caption{The branching ratio of various possible decay modes of the scalar $s_5$, assuming $\Lambda=500$ GeV.    }
  \label{figh5br}
	\end{figure}

\begin{figure}[H]
	\centering
 \begin{subfigure}[]{0.327\linewidth}
 \includegraphics[width=\linewidth]{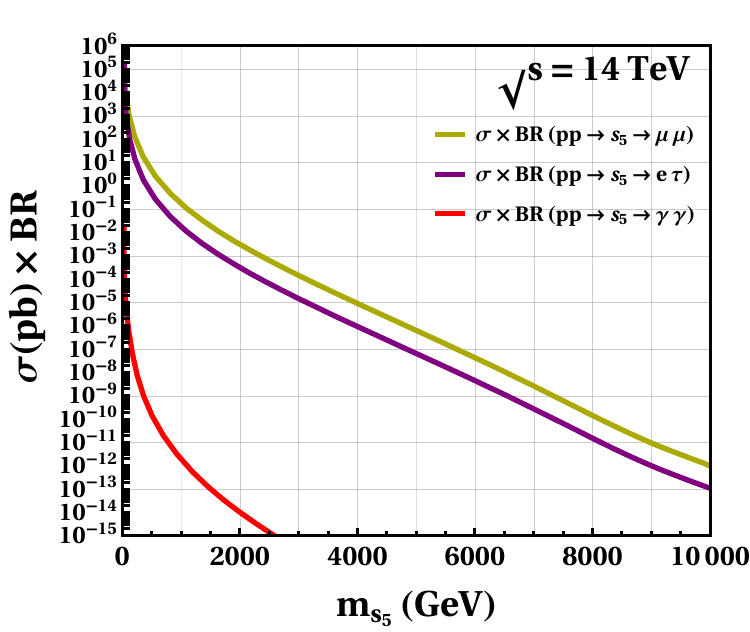}
 \caption{}
         \label{figh5a}
 \end{subfigure} 
 \begin{subfigure}[]{0.327\linewidth}
    \includegraphics[width=\linewidth]{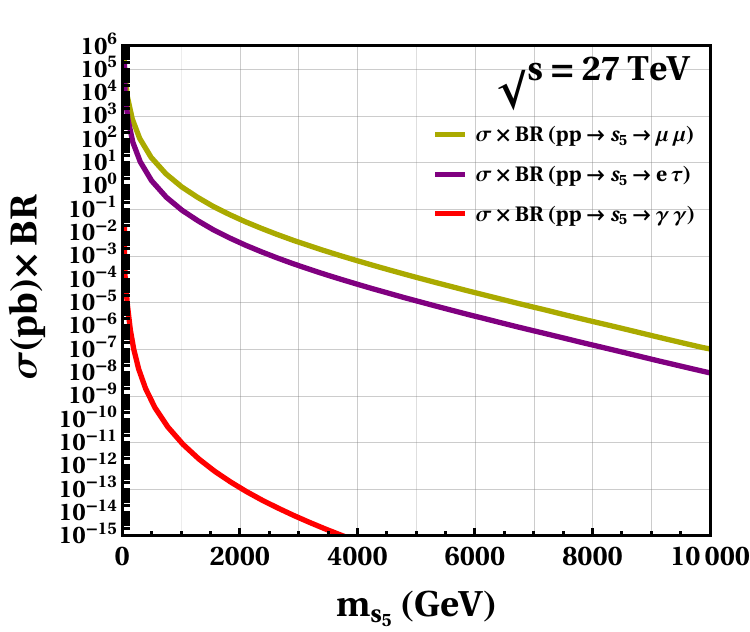}
    \caption{}
         \label{figh5b}	
\end{subfigure}
\begin{subfigure}[]{0.327\linewidth}
    \includegraphics[width=\linewidth]{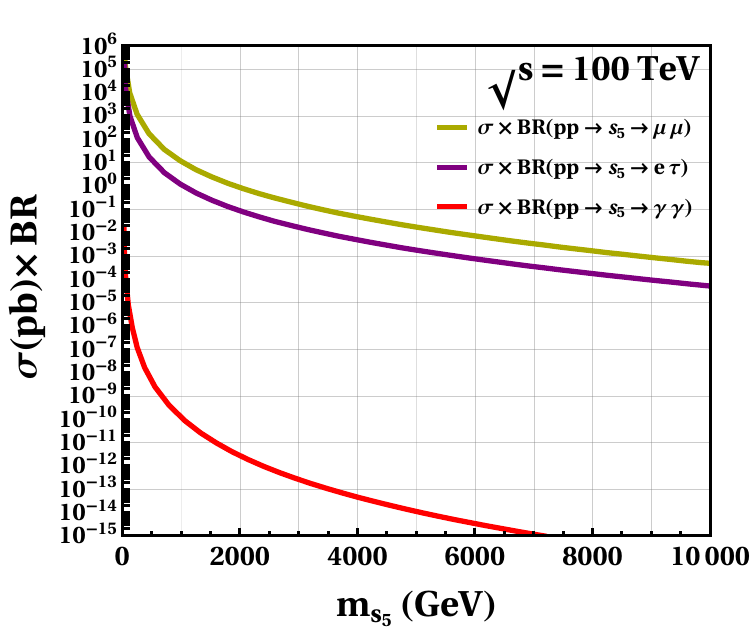}
    \caption{}
         \label{figh5c}	
\end{subfigure}
 \caption{Cross-section ($\sigma \times BR$) for the processes  $pp \rightarrow s_5 \rightarrow \mu \mu$, $pp \rightarrow s_5 \rightarrow e \tau$, and $pp \rightarrow s_5 \rightarrow \gamma \gamma$ as functions of the scalar mass $m_{s_5}$, are shown in figure \ref{figh5a} for the 14 TeV HL-LHC, figure \ref{figh5b} for the 27 TeV HE-LHC, and in figure \ref{figh5c} for a 100 TeV collider, where we have set the scale $\Lambda=500$ GeV.   }
  \label{figh5sigma}
	\end{figure}

We now examine the signatures of the scalar $s_5$. Its potential decay modes include $s \bar{s}$, $t \bar{u}$ as well as the leptonic channels $\mu \bar{\mu}$, and $ \bar{e} \tau$.  Additionally, there are loop-induced modes $gg$ and $\gamma \gamma$, which are significantly suppressed  in comparison. The branching fractions of these decay channels as a function of the mass of the scalar $s_5$, are illustrated in the figure \ref{figh5br}. The production cross-sections of the scalar $s_5$,  as a function of the scalar mass $m_{s_5}$, are shown in figure \ref{figh5sigma}   at the HL-LHC, HE-LHC, and a 100 TeV collider.

The benchmark predictions for the production cross-sections at the HL-LHC, HE-LHC, and a 100 TeV collider for the heavy and light scalar mass $m_{s_5}$  are listed in the tables \ref{tab:benchh5_heavy} and \ref{tab:benchh5_light}, respectively, where we set the scale $\Lambda=500$ GeV. For heavy scalar mass $m_{s_5}$, the inclusive channels $\mu \bar{\mu}$ and $ \bar{e} \tau$ are accessible to HL-LHC, HE-LHC, as well as a 100 TeV collider. In contrast, the di-photon $\gamma \gamma$ channel remains insensitive to the HL-LHC, HE-LHC, and a 100 TeV collider, regardless of whether the scalar $s_5$ is  heavy or light.


\begin{table}[H]
\setlength{\tabcolsep}{6pt} 
\renewcommand{\arraystretch}{1.2} 
\centering
\begin{tabular}{l|cc|cc|cc}
\toprule
 & \multicolumn{2}{c|}{HL-LHC [14 TeV, $3~\iab$] } & \multicolumn{2}{c|}{HE-LHC [27 TeV, $15~\iab$]} & \multicolumn{2}{c}{100 TeV, $30~\iab$} \\
$m_{s_5}$~[GeV] &  500 &  1000 &  500 &  1000 &  500 &  1000 \\
\midrule
$\mu \mu$ [pb]     & \fbox{$4.5$}   & \fbox{$0.2$}      & \fbox{$16.4$}     & \fbox{$1.0$}  & \fbox{$124.9$} & \fbox{$11.9$}  \\
$e \tau$ [pb]     & \fbox{$0.43$}   & \fbox{$1.8\e{-2}$ }     & \fbox{$1.6$}     & \fbox{$0.1$}  & \fbox{$12$} & \fbox{$1.2$}  \\
$\gamma \gamma$ [pb] & $1\e{-10}$  & $2\e{-12}$  & $5\e{-10}$& $1\e{-11}$  & $4\e{-9}$ & $1\e{-10}$  \\
\bottomrule
\end{tabular}
\caption{Benchmark ($\sigma \times BR$) for higher values of the scalar mass $m_{s_5}$ at $\Lambda = 500$ GeV, where $k_{1-6}= 1.1 \times 10^{6}$ for $m_{s_5}= 500$ GeV, and $k_{1-6}= 4.6 \times 10^{6}$ for $m_{s_5}= 1000$ GeV.}
\label{tab:benchh5_heavy}
\end{table}

The scalar $s_5$ can have a minimum mass of approximately $36$ GeV, when the proportionality constant is set to its minimum allowed value, $k_{1-6}=6 \times 10^3$ at the scale $\Lambda=500$ GeV. Therefore, we do not show the benchmark predictions for the masses of scalar $s_5$ below  $36$ GeV.

\begin{table}[H]
\setlength{\tabcolsep}{6pt} 
\renewcommand{\arraystretch}{1} 
\centering
\begin{tabular}{l|c|c|c}
\toprule
 & HL-LHC [14 TeV, $3~\iab$] & HE-LHC [27 TeV, $15~\iab$] & 100 TeV, $30~\iab$ \\
$m_{s_5}$~[GeV] & 60 & 60 & 60 \\
\midrule
$\gamma \gamma$ [pb] & $9\e{-6}$ & $2\e{-5}$ & $7\e{-5}$ \\
\bottomrule
\end{tabular}
\caption{Benchmark ($\sigma \times BR$) for low values of the scalar mass $m_{s_5}$ at $\Lambda = 500$ GeV, where $k_{1-6}= 1.6 \times 10^{4}$ for $m_{s_5}= 60$ GeV.}
\label{tab:benchh5_light}
\end{table}

\subsubsection{Scalar $s_6$}
\begin{figure}[H]
\centering
    \includegraphics[width=0.45\linewidth]{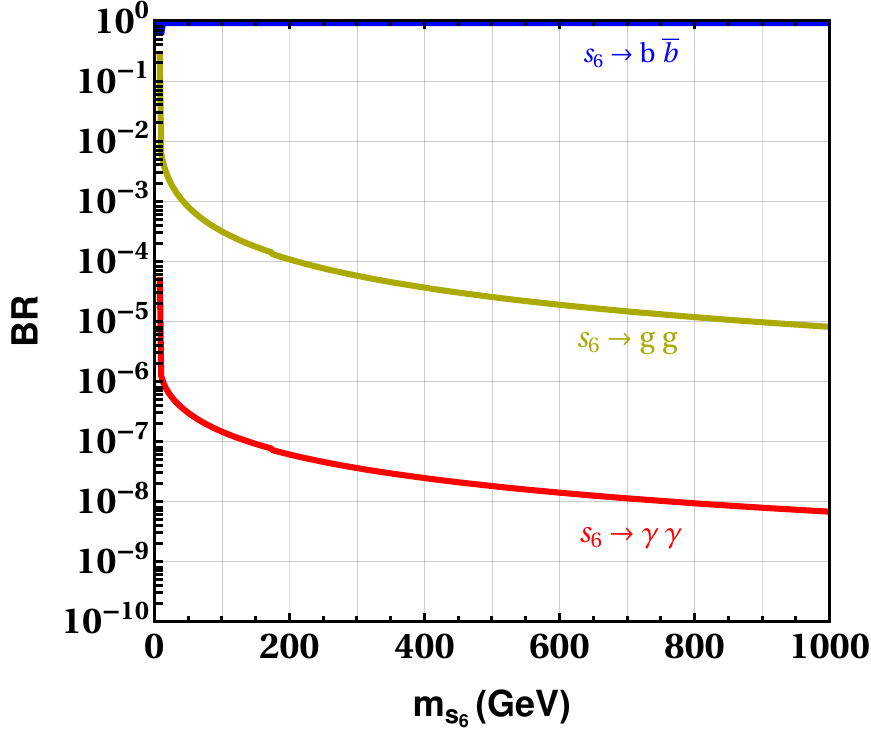}
 \caption{The branching ratio of various possible decay modes of the scalar $s_6$, assuming $\Lambda=500$ GeV.    }
  \label{figh6br}
	\end{figure}

\begin{figure}[H]
	\centering
 \begin{subfigure}[]{0.327\linewidth}
 \includegraphics[width=\linewidth]{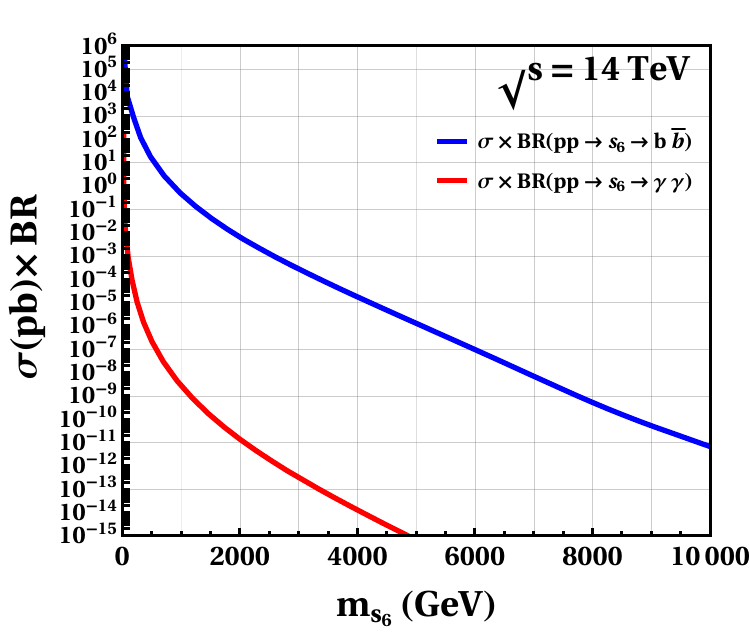}
 \caption{}
         \label{figh6a}
 \end{subfigure} 
 \begin{subfigure}[]{0.327\linewidth}
    \includegraphics[width=\linewidth]{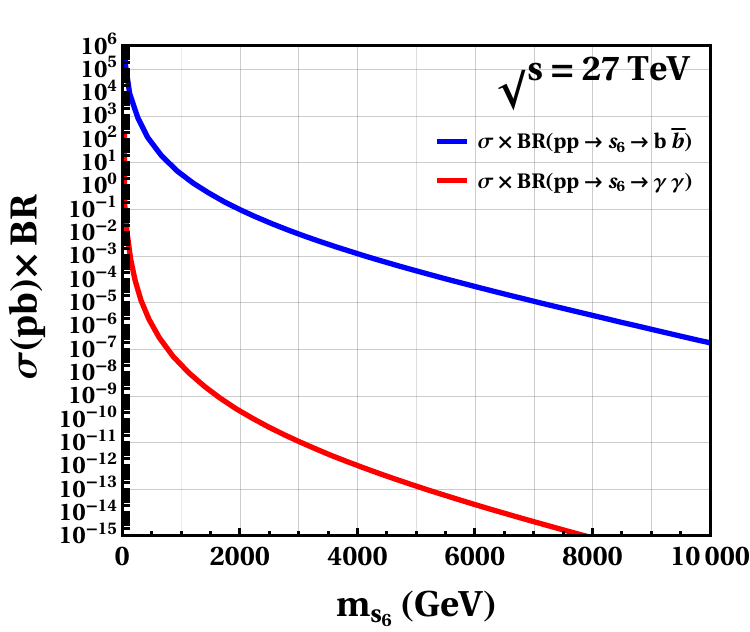}
    \caption{}
         \label{figh6b}	
\end{subfigure}
\begin{subfigure}[]{0.327\linewidth}
    \includegraphics[width=\linewidth]{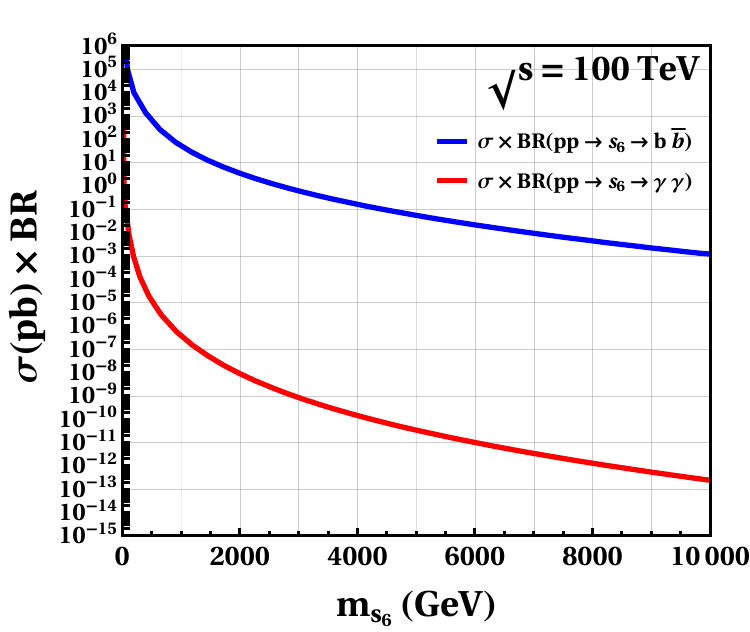}
    \caption{}
         \label{figh6c}	
\end{subfigure}
 \caption{Cross-section ($\sigma \times BR$) for the processes  $pp \rightarrow s_6 \rightarrow b \overline{b}$ and $pp \rightarrow s_6 \rightarrow \gamma \gamma$ as functions of the scalar mass $m_{s_6}$, are shown in figure \ref{figh6a} for the 14 TeV HL-LHC, figure \ref{figh6b} for the 27 TeV HE-LHC, and in figure \ref{figh6c} for a 100 TeV collider, where we have set the scale $\Lambda=500$ GeV.  }
  \label{figh6sigma}
	\end{figure}

The scalar $s_6$ only couples to   $b \bar{b}$ pair. The branching fractions of the scalar $s_6$ to   $b \bar{b}$   and  the loop-induced $gg$ and $\gamma \gamma$ channels, are shown in figure \ref{figh6br}. It is evident that the  decay into $b \bar{b}$ pair dominates.   The production cross-sections of the scalar $s_6$ for the inclusive channels $b \bar{b}$ and $\gamma \gamma$ at the HL-LHC, HE-LHC, and a 100 TeV collider, are shown in  figure \ref{figh6sigma}. \\

The benchmark predictions for the production cross-sections of the scalar $s_6$ at the scale $\Lambda = 500$ GeV, are given in table \ref{tab:benchh6_heavy}. For a heavy scalar mass $m_{s_6}$, the inclusive $b \bar{b}$  channel is accessible at the HL-LHC, HE-LHC, and a 100 TeV collider.  The di-photon $\gamma \gamma$ channel remains beyond the reach of all the three future colliders.

\begin{table}[H]
\setlength{\tabcolsep}{6pt} 
\renewcommand{\arraystretch}{1.2} 
\centering
\begin{tabular}{l|c|c|c}
\toprule
 & HL-LHC [14 TeV, $3~\iab$] & HE-LHC [27 TeV, $15~\iab$] & 100 TeV, $30~\iab$ \\
$m_{s_6}$~[GeV] & 1000 & 1000 & 1000 \\
\midrule
$b \bar{b}$ [pb]     & \fbox{$0.44$} & \fbox{$3.2$} & \fbox{$56.5$} \\
$\gamma \gamma$ [pb] & $3\e{-9}$     & $2\e{-8}$    & $4\e{-7}$     \\
\bottomrule
\end{tabular}

\caption{Benchmark ($\sigma \times BR$) for higher values of the scalar mass $m_{s_6}$ at $\Lambda = 500$ GeV, where $k_{1-6}= 1.1 \times 10^{4}$ for $m_{s_6}= 1000$ GeV.}
\label{tab:benchh6_heavy}
\end{table}

The scalar $s_6$ can have a minimum mass of approximately $7.24\times 10^2$ GeV, when the proportionality constant in equation \ref{sig_r} is set to its minimum allowed value, $k_{1-6}=6 \times 10^3$ at the scale $\Lambda=500$ GeV. Therefore, the benchmark predictions for the masses of scalar $s_6$ below  $7.24\times 10^2$ GeV are excluded for the scale $\Lambda=500$ GeV.

\subsubsection{Scalar $s_7$}
The  scalar $s_7$ couples to $\nu \nu$ pair, and its coupling to other fermions effectively negligible.    Therefore, it does not show any appreciable signature in collider physics.  However, $s_7$ will uniquely contribute to $\nu\nu \rightarrow \nu \nu$ scattering.  Such an investigation is beyond the scope of this paper.

\begin{table}[H]
    \setlength{\tabcolsep}{6pt}
    \renewcommand{\arraystretch}{1.34}
    \centering
    \begin{tabular}{|c|c|c|}
        \hline
        \textbf{Scalar} & \textbf{Final states} & \textbf{Masses [GeV]} \\
        \hline
        $s_1$ & $e e$         & 500, 1000 \\ \hline
        $s_2$ & $\tau \tau$   & 500, 1000 \\
              & $\mu \tau$    & 500, 1000 \\ \hline
        $s_4$ & $e \mu$       & 500, 1000 \\ \hline
        $s_5$ & $\mu \mu$     & 500, 1000 \\
              & $e \tau$      & 500, 1000 \\ \hline
        $s_6$ & $b \bar{b}$   & 1000 \\
        \hline
    \end{tabular}
    \caption{Summary of accessible inclusive production channels of scalars $s_i$ in the SHVM with benchmark masses. }
    \label{tab:summ_scalars}
\end{table}

Table \ref{tab:summ_scalars} summarizes the accessible inclusive production channels of the scalars $s_i$ in the SHVM, along with their corresponding benchmark mass values. As evident from the table, each scalar $s_i$ exhibits a distinct final-state signature, which is non-overlapping with the signatures of the other scalars. This feature of mutually exclusive final states sets the SHVM apart from other models with extended scalar sectors. Indeed, no known model in the literature simultaneously accommodates all the scalar signatures listed in Table \ref{tab:summ_scalars}, highlighting the unique predictive structure of the SHVM.
Such model-specific signatures offer a clear path for experimental discrimination and could be systematically probed at the HL-LHC, HE-LHC, and a future 100 TeV collider providing a robust strategy for distinguishing the SHVM from other scalar sector extensions.

\section{ 95.4 GeV excess}
\label{anomaly}
We now discuss a low-mass di-photon excess reported by the CMS collaboration at $m_{\gamma\gamma}$ = 95.4 GeV with a local significance of $2.9\sigma$ \cite{cms95}.  This excess is also supported by the similar results reported by the ATLAS collaboration  with a local significance of $1.7\sigma$  \cite{atlas95}.  The combined signal strength can be written as \cite{Biekotter:2023oen},

\begin{align}
\mu_{\gamma\gamma}^{exp}= & \mu_{\gamma\gamma}^{ATLAS+CMS}=\frac{\sigma(pp\to\phi\to \gamma\gamma)}{\sigma_{SM}(pp\to h^{SM}_{95.4}\to \gamma\gamma)}=0.24^{+0.09}_{-0.08},
\end{align}
 with a local significance of $ 3.1 \sigma$. We notice that  $\phi$  represents  a non-SM scalar with a mass of 95.4 GeV, and the scalar $h^{SM}_{95.4}$ shows a SM-like Higgs with the same mass.  There are numerous explanation to this observation \cite{Biekotter:2023oen}-\cite{Arhrib:2024wjj}.  

 We also note that there is another local excess in the same mass region  in the $e^+ e^- \to Z(\phi\to b\bar{b})$ searches at LEP \cite{LEPWorkingGroupforHiggsbosonsearches:2003ing},
\begin{align}
\mu_{b\bar{b}}^{exp}= & 0.117\pm 0.057.
\end{align}
This claim is confronted  in reference \cite{Janot:2024ryq}, which strongly disfavours this excess.  In the SHVM as well, this excess is not present since the pseudoscalar $a_3$ cannot decay to  $b\bar{b}$ mode.  Thus, if the 95.4 excess continues to show up in future runs of the LHC, and it only exits in the di-photon channel, the SHVM will be an important  framework to address this excess.

\begin{figure}[H]
	\centering
 \includegraphics[width=0.45\linewidth]{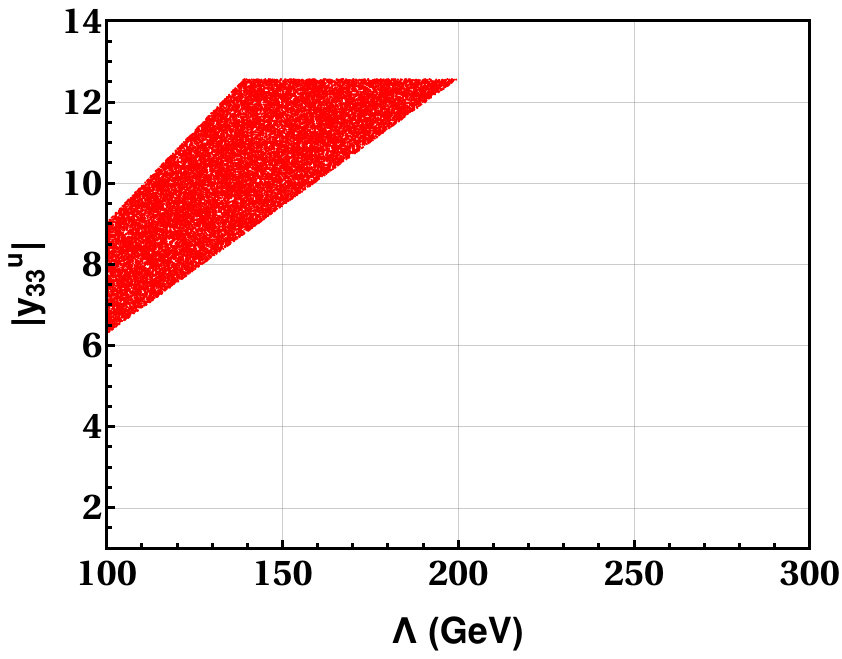}
 \caption{Allowed values of $y_{33}^u$ and $\Lambda$ (red colored region) for $a_3$  to address the di-photon excess at $m_{\gamma\gamma}$ = 95.4 GeV.  }
  \label{fig_excess}
	\end{figure}

In the SHVM, this excess can be addressed through the pseudoscalar $a_3$, which is insensitive to the flavor bounds.  There are only two free parameters $|y_{33}^u|$ and $\Lambda$  in producing the 95.4 GeV excess.  This is because  the coupling of  the pseudoscalar $a_3$ to   $t \bar{t}$ pair is  independent of the parameter $\epsilon_3$. We assume that the range of the parameter $|y_{33}^u|$ to be $1-4\pi$, which is allowed by perturbation theory.   Moreover, we are assuming that the underlying ultraviolet completion of the SHVM lies in a strongly interacting dynamics as discussed in the appendix.  Therefore, even a large value of $|y_{33}^u|$ is allowed.   We show the variation of the $|y_{33}^u|$ and $\Lambda$, which can produce the 95.4 excess, in figure \ref{fig_excess}.

Thus, we observe that the SHVM without large FCNC effects  has potential to reveal itself  even closer to the electroweak scale by accommodating an anomalous mass state such as 95.4 GeV excess.  Hence, the SHVM may play an interesting role in addressing several anomalies given in table  \ref{tab:anomalies} if they continue to show up in future runs of the LHC.  However, such an investigation is beyond the scope of this work, and will be discussed in an upcoming work.

\section{Neutrinic dark matter} 
\label{neutrinic_dm}
In this section, we show that the axial degrees of freedom of the singlet field $\chi_7$ becomes a new class of neutrino-philic dark matter, since   $a_7$ decays only to a pair of neutrinos.      The main difference,  which keeps  the  neutrino-philic-neutrinic dark matter apart from standard axion-like particle dark matter,  is that the neutrinic   dark matter  does not couple to the electromagnetic field at leading order, and the decay $a_7 \rightarrow \gamma \gamma$ occurs at two-loop, which is effectively negligible.  

The mass of $a_7$ is given by
\begin{align}
\label{a7_mass}
m_{a_7}^2 = & 4 \rho_7^2 -  9 \sqrt{2} \epsilon_7 \Lambda \sigma_7.
\end{align}
For  $a_7$ to be a neutrinic  dark matter, we can use the misalignment mechanism  \cite{Preskill:1982cy,Abbott:1982af,Dine:1982ah} where the true vacuum and initial value of the field $a_7$  during inflation could be misaligned. This is followed by the  rolling down of  $a_7$ fields to the true vacuum resulting in cold dark matter density of coherent oscillation.  

The equation of motion of a scalar field  in the expanding universe is, 
\begin{equation}
\label{eq_mot}
   \ddot{\phi}+3H \dot{\phi} + m_\phi^2 \phi \approx 0,
\end{equation}
where $\phi \equiv  a_7 $.  

The solution of  equation \ref{eq_mot} is given by   $\phi(t)=  \phi_0  2^{1\over4} J_{1\over4}(m_\phi t)/(m_\phi t)^{\frac{1}{ 4}}$  where $\phi_0$ is the initial value of the field   $\phi \equiv  a_7 $.  The evolution of the energy density $\rho_\phi = {1\over2}(\dot\phi^2+m_\phi^2 \phi^2) $ at later time ($m_\phi t\to \infty$)  turns out to be  $\rho_\phi \approx m^2_\phi \phi_0^2 \sqrt{2}\Gamma(5/4)^2/\pi (m_\phi t)^{3/2}$ \cite{Abbas:2023ion}.  Using  the dark matter density, $\rho_\phi = 0.24\, {\rm eV}^4$ at the matter-radiation equality time $t_{eq}$, that is, $m_\phi t_{eq} \approx 2 \times 10^{27} (m_\phi/{\rm eV})$, and equating it with the $\rho_\phi \approx m^2_\phi \phi_0^2 \sqrt{2}\Gamma(5/4)^2/\pi (m_\phi t)^{3/2}$,  we obtain the mass of dark matter particle which is,
\begin{equation}
\label{mphi2}
  m_\phi =3.4\times 10^{-3} {\rm eV} \left( \frac{10^{12} {\rm GeV}}{ \phi_0} \right)^4.
\end{equation}
Now, equating the above mass with the mass given in equation \ref{a7_mass}, we get
\begin{align}
    \sigma_7 = \dfrac{-7.75 \times 10^{71} + 0.32 \phi_0^8 \rho_7^2 }{\epsilon_7 \Lambda \phi_0^8}.
\end{align}
Substituting the above result back into the equation \ref{a7_mass}, we obtain the mass of neutrinic dark matter,
\begin{align}
    m_{a_7} = 3.14 \times 10^{-3} {\rm eV}  \left( \frac{10^{12} {\rm GeV}}{ \phi_0} \right)^4.
\end{align}
The cold dark matter particle is bounded from below to be $10^{-21}$ eV \cite{Cirelli:2024ssz}.  We consider the scenario where $m_{a_7} < 2 m_{\nu_1}$, where  $m_{\nu_1} = 2.67 \times 10^{-4}$ eV is the lightest neutrino mass given in equation \ref{neut_mass}.  In this case, the decay $a_7 \rightarrow \nu \nu $ is forbidden and the scale $\Lambda$ can be close to the electroweak scale.

\begin{figure}[H]
	\centering
 \includegraphics[width=0.54\linewidth]{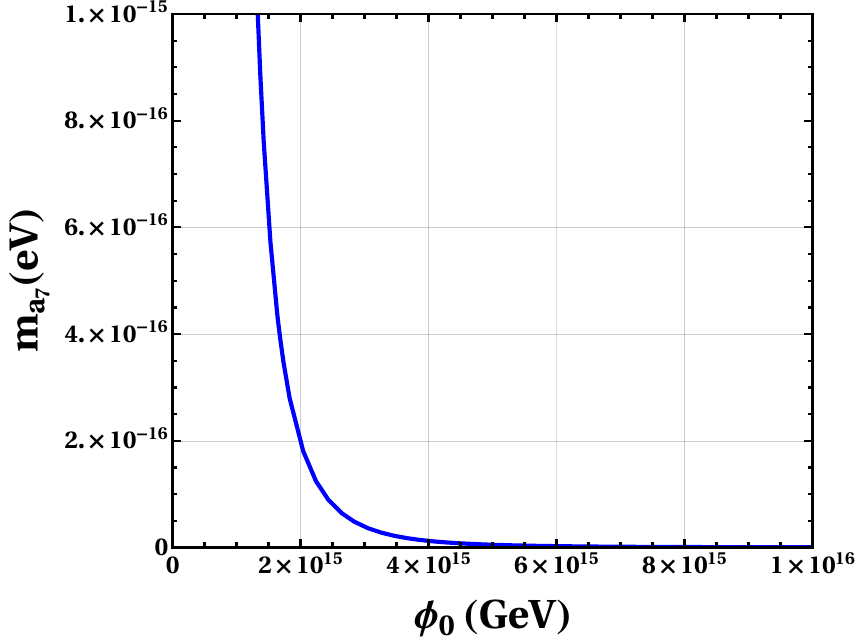}
 \caption{The allowed mass range of the neutrinic dark matter for the scenario $m_{a_7} < 2 m_{\nu_1}$.  }
  \label{fig_dm}
	\end{figure}

In figure \ref{fig_dm}, we show the variation of the parameter $\phi_0$ relative to the mass of the neutrinic dark matter when  $m_{a_7} < 2 m_{\nu_1}$.  In this scenario, the neutrinic dark matter can reveal itself via the Higgs-portal through the term $\varphi^\dagger \varphi \chi_7^* \chi_7$ or through the term $\chi_r^*\chi_r  \chi_7^* \chi_7$.  However, such an investigation is beyond the scope of  this paper.

\section{Summary}
\label{sum}
The SHVM presents an unconventional framework for addressing the flavor problem. As established in Ref.~\cite{Abbas:2025ser}, the SHVM emerges as a low-energy effective limit of the DTC paradigm. Within the DTC framework, the VEVs of the gauge, singlet scalar fields $\chi_r$ correspond to multi-fermion bound states, a point that is briefly discussed in the appendix of this paper.

In the present work, we investigate  a version of the SHVM in which protection against large FCNC effects is built into the framework. This is accomplished by incorporating hierarchical soft symmetry-breaking parameters, $\rho_r$ and $\sigma_r$. Within this setup, we investigate the collider phenomenology of the model and identify the mutually exclusive, defining signatures associated with each scalar $s_r$. 
  
The scalars $s_r$ exhibit a variety of rich and distinct collider signatures. For instance, the scalar $s_1$ can be probed via the $ee$ channel at the HL-LHC, HE-LHC, and a future 100 TeV collider for heavy mass benchmarks. Similarly, $s_2$ is accessible through the $\tau\tau$ and $\mu \tau$  final states at all three collider platforms. The scalar $s_4$ can be searched for in the  $e\mu$ channel, while $s_5$ is accessible in both the $\mu\mu$ and $e \tau$ modes across all considered colliders. Lastly, the scalar $s_6$ can be probed through the $b\bar{b}$ final state, again at the HL-LHC, HE-LHC, and 100 TeV collider. These distinctive signatures serve as the unique fingerprint of the SHVM, providing a clear means to differentiate it from other models with extended scalar sectors. Additionally, we also found that pseudoscalar $a_3$ can explain the observed 95.4 GeV di-photon excess.

Furthermore, the pseudoscalar $a_7$ gives rise to an entirely new class of dark matter, which we refer to as neutrinic dark matter, due to its unique interactions exclusively with neutrino pairs. The phenomenology of neutrinic dark matter remains an open avenue and will be the subject of future investigation.

\section*{Acknowledgement}
We thank N. R. S. Chundawat for collaboration during the very early stages of this work. This paper is dedicated to the memory of Ashutosh Kumar Alok, a remarkable human being whose scientific insight and dedication greatly influenced this research. We also thank Vartika Singh for carefully reading the manuscript. This work is supported by the  Council of Science and Technology,  Govt. of Uttar Pradesh,  India through the  project ``   A new paradigm for flavor problem "  no.   CST/D-1301,  and Anusandhan National Research Foundation (Science and Engineering Research Board) , Department of Science and Technology, Government of India through the project `` Higgs Physics within and beyond the Standard Model" no. CRG/2022/003237.  NS acknowledges the support through the INSPIRE fellowship by the Department of Science and Technology, Government of India.

 \section*{Data availability statements}
 The used data is explicitly quoted in the manuscript itself, and there is no need to deposit it separately.  The data of the plots presented in this work  as well as the  mathematica codes will be available on request.

\section*{Appendix}

\section*{Benchmark points for the Yukawa couplings}
We reproduce the fermion masses using the following values of the fermion masses at $ 1$ TeV \cite{Xing:2007fb},
\begin{eqnarray}
\{m_t, m_c, m_u\} &\simeq& \{150.7 \pm 3.4,~ 0.532^{+0.074}_{-0.073},~ (1.10^{+0.43}_{-0.37}) \times 10^{-3}\}~{\rm GeV}, \nonumber \\
\{m_b, m_s, m_d\} &\simeq& \{2.43\pm 0.08,~ 4.7^{+1.4}_{-1.3} \times 10^{-2},~ 2.50^{+1.08}_{-1.03} \times 10^{-3}\}~{\rm GeV},
\nonumber \\
\{m_\tau, m_\mu, m_e\} &\simeq& \{1.78\pm 0.2,~ 0.105^{+9.4 \times 10^{-9}}_{-9.3 \times 10^{-9}},~ 4.96\pm 0.00000043 \times 10^{-4}\}~{\rm GeV}.
\end{eqnarray}

The magnitudes and phases  of the CKM mixing elements are \cite{Zyla:2021},
\bea
|V_{ud}| &=& 0.97370 \pm 0.00014,  |V_{cb}| = 0.0410 \pm 0.0014, |V_{ub}| = 0.00382 \pm 0.00024, \\ \nonumber
\sin 2 \beta &=& 0.699 \pm 0.017, ~ \alpha = (84.9^{+5.1}_{-4.5})^\circ,~  \gamma = (72.1^{+4.1}_{-4.5})^\circ, \delta = 1.196^{+0.045}_{-0.043}
\eea

We conduct a $\chi^2$ fit to the masses of quarks and charged leptons, as well as the parameters describing quark mixing, by defining,
\bea
\chi^2 &=& \dfrac{(m_q - m_q^{\rm{model}} )^2}{\sigma_{m_q}^2}+  \dfrac{(m_\ell - m_\ell^{\rm{model}} )^2}{\sigma_{m_\ell}^2}  + \dfrac{(\sin \theta_{ij} - \sin \theta_{ij}^{\rm{model}} )^2}{\sigma_{\sin \theta_{ij}}^2} + \dfrac{(\sin 2 \beta  - \sin 2 \beta^{\rm{model}} )^2}{\sigma_{\sin2\beta}^2} \nonumber \\ &+& \dfrac{( \alpha  - \alpha^{\rm{model}} )^2}{(\sigma_{\alpha})^2}
+ \dfrac{( \gamma   - \gamma^{\rm{model}} )^2}{(\sigma_{\gamma})^2},
\eea
where $q=\{u,d,c,s,t,b\}$, $\ell=\{e,\mu,\tau\}$, and $i,j=1,2,3$.

The dimensionless coefficients $y_{ij}^{u,d,\ell,\nu}= |y_{ij}^{u,d,\ell,\nu}| e^{i \phi_{ij}^{q,\ell,\nu}}$  are scanned with $|y_{ij}^{u,d,\ell, \nu}| \in [0.1, 4 \pi]$ and $ \phi_{ij}^{q,\ell,\nu} \in [0,2\pi]$. The fit results  are,
\begin{equation*}
y^u_{ij} = \begin{pmatrix}
-1.966 + 0.368i & 0 & 0.044 - 11.583i \\
0 & -0.010 + 1.000i & -0.384 - 11.518i \\
0 & 0 & -0.909 - 0.417i \\
\end{pmatrix},  
\end{equation*}

\begin{equation*}
y^d_{ij} = \begin{pmatrix}
3.482 + 2.922i & -0.938 - 0.347i & 0 \\
0 & -0.917 - 0.392i & 0 \\
0 & 0 & -0.934 - 2.412i \\
\end{pmatrix}.
\end{equation*}
We obtain $\delta_{\rm CP}^q \approx 1.144$ for  Dirac CP phase.

\begin{equation*}
y^\ell_{ij} = \begin{pmatrix}
 0.902\, -0.093 i & 0.523\, -0.887 i & 0.356\, -1.041 i \\
 0 & 0.295\, +2.341 i & 0.422\, +1.016 i \\
 0 & 0 & 3.173\, + 5.834 \times 10^{-9} i \\
\end{pmatrix},
\end{equation*}

The neutrino couplings for normal mass ordering are,
\begin{equation*}
y^\nu_{ij} = \begin{pmatrix}
0.6\, -0.88 i & 0.87\, -0.41 i & 1.5\, -0.00004 i \\
0 & 0.96\, +0.12 i & -0.52-1.41 i \\
0 & -1.43+0.28 i & 1.5\, +0.00004 i
\end{pmatrix}, 
\end{equation*}
and the leptonic Dirac $CP$ phase  turns out to be $\delta_{\rm CP}^\ell  \approx  3.14$, and $\chi^2 = 5.71.$

\subsection*{Bi-unitary transformation matrices }

\begin{equation*}
U_u = 
\begin{pmatrix}
-0.188 + 0.982i & 5.706 \times 10^{-8} - 2.983 \times 10^{-7}i & -4.943 \times 10^{-9} + 2.584 \times 10^{-8}i \\
-3.034 \times 10^{-7} + 1.304 \times 10^{-8}i & -0.999 + 0.043i & -1.422 \times 10^{-4} + 6.110 \times 10^{-6}i \\
-2.635 \times 10^{-8}  & -1.423 \times 10^{-4}  & 1 \\
\end{pmatrix}
\label{bi-uni-up1_shvm2}
\end{equation*}

\begin{equation*}
V_u = 
\begin{pmatrix}
-0.414 - 0.910i & 6.052 \times 10^{-5} + 1.331 \times 10^{-4}i & 1.493 \times 10^{-3} + 3.283 \times 10^{-3}i \\
-2.820 \times 10^{-10} - 5.636 \times 10^{-10}i & -0.447 - 0.894i & 1.812 \times 10^{-2} + 3.622 \times 10^{-2}i \\
3.610 \times 10^{-3}  & 4.050 \times 10^{-2}  & 0.999 \\
\end{pmatrix}
\label{bi-uni-up2_shvm2}
\end{equation*}

\begin{equation*}
U_d =
\begin{pmatrix}
-0.941 + 0.337i & -0.011 + 0.004i & 0  \\
-0.012  & 0.999  & 0  \\
0  & 0  & 1 \\
\end{pmatrix}
\label{bi-uni-dw1_shvm2}
\end{equation*}

\begin{equation*}
V_d =
\begin{pmatrix}
-0.974 + 0.049i & 0.221 - 0.011i & 0  \\
0.221  & 0.975  & 0  \\
0  & 0  & 1  \\
\end{pmatrix}
\label{bi-uni-dw2_shvm2}
\end{equation*}

\begin{equation*}
U_\ell =
\begin{pmatrix}
-1.457 \times 10^{-1} - 9.893 \times 10^{-1}i & -1.572 \times 10^{-4} + 4.160 \times 10^{-4}i & 3.416 \times 10^{-6} - 7.407 \times 10^{-6}i \\
 -3.177 \times 10^{-4} + 3.113 \times 10^{-4}i & -9.642 \times 10^{-1} + 2.645 \times 10^{-1}i & 1.989 \times 10^{-2} - 5.455 \times 10^{-3}i \\
 1.188 \times 10^{-6}  & 2.062 \times 10^{-2}  & 9.998 \times 10^{-1}  \\
 \end{pmatrix}
 \label{bi-uni-lep1}
\end{equation*}

\begin{equation*}
V_\ell =
\begin{pmatrix}
 -2.455 \times 10^{-1} - 9.644 \times 10^{-1}i & -2.395 \times 10^{-2} + 9.067 \times 10^{-2}i & 9.415 \times 10^{-3} - 2.756 \times 10^{-2}i \\
 -7.731 \times 10^{-2} - 6.040 \times 10^{-2}i & -3.598 \times 10^{-1} - 8.677 \times 10^{-1}i & 1.260 \times 10^{-1} + 3.034 \times 10^{-1}i \\
 4.024 \times 10^{-3}  & 3.298 \times 10^{-1}  & 9.440 \times 10^{-1}  \\
\end{pmatrix}
\label{bi-uni-lep2}
\end{equation*}

\subsection*{Dark-technicolor paradigm}
We now discuss the dark-technicolor  paradigm, which gives rise to the SHVM \cite{Abbas:2020frs}.  The dark-technicolor paradigm is based on the symmetry $\mathcal{S} =  SU(3)_c \times SU(2)_L \times U(1)_Y \times SU(\rm{N}_{\rm{TC}}) \times SU(\rm{N}_{\rm{DTC}}) \times SU(\rm{N}_{\rm{F}})  $, where TC stands for technicolor, DTC is for the  dark-technicolor, and $\rm F$ is a strong QCD-like dynamics of vector-like fermions.  

We have the following transformations of TC fermions under the symmetry    $\mathcal{S} $ \cite{Abbas:2020frs},
\begin{eqnarray}
{T_L}^i  &\equiv&   \begin{pmatrix}
T_i  \\
B_i
\end{pmatrix}_L:(1,2,0, \rm{N}_{ TC},1,1),  
T_{i, R} : (1,1,1, \rm{N}_{ TC},1,1), \\ \nonumber
 B_{i, R}  &:& (1,1,-1, \rm{N}_{ TC},1,1), 
\end{eqnarray}
where the fermions $T_i$ are  assigned electric charges $+\frac{1}{2}$,  and that of the fermions  $B_i$ is $-\frac{1}{2}$. 

The $\rm DTC$ vector-like fermions  transform under the symmetry $\mathcal{S} $ as,
\begin{eqnarray}
 D  &\equiv& C_{i,  L,R}: (1,1, 1,1, \rm{N}_{\rm DTC},1),~ S_{i, L,R} : (1,1,-1,1, \rm{N}_{\rm DTC},1), 
\end{eqnarray} 
where  $+\frac{1}{2}$ is the electric charge for the quark $\mathcal C$, and that of the  $\mathcal S$ is  $-\frac{1}{2}$.   

The transformation of the vector-like fermions of  the symmetry $ SU(\rm N_{\rm F})$ is,
\begin{eqnarray}
F_{L,R} &\equiv &U_{L,R}^i \equiv  (3,1,\dfrac{4}{3},1,1, \rm{N}_{\rm F}),
D_{L,R}^{i} \equiv   (3,1,-\dfrac{2}{3},1,1,\rm{N}_{\rm F}),  \\ \nonumber 
N_{L,R}^i &\equiv&   (1,1,0,1,1,\rm{N}_{\rm F}), 
E_{L,R}^{i} \equiv   (1,1,-2,1,1,\rm{N}_{\rm F}).
\end{eqnarray}

There exist three global axial $U(1)_A^{\rm TC, DTC,  F}$ symmetries  in the dark-technicolor paradigm, which  are broken by instanton effects as $ U(1)_A^{\rm TC, DTC, F} \rightarrow \mathcal{Z}_{2 \rm K_{\rm TC, DTC, F}}$ providing a generic $\mathcal{Z}_{\rm N} \times \mathcal{Z}_{\rm M} \times \mathcal{Z}_{\rm P}$ flavor symmetry, where $\rm N= 2 \rm K_{\rm TC}$, $\rm M= 2 \rm K_{\rm DTC}$ and $\rm P= 2 \rm K_{\rm F}$, and  $\rm  \rm K_{\rm TC}$, $\rm  \rm K_{\rm DTC}$ and $\rm  \rm K_{\rm F}$ represent the number of massless flavors in the fundamental representation of the gauge group  $ SU(\rm N_{\rm TC, DTC,  F})$  \cite{Harari:1981bs}. 

\begin{figure}[h]
	\centering
 \includegraphics[width=\linewidth]{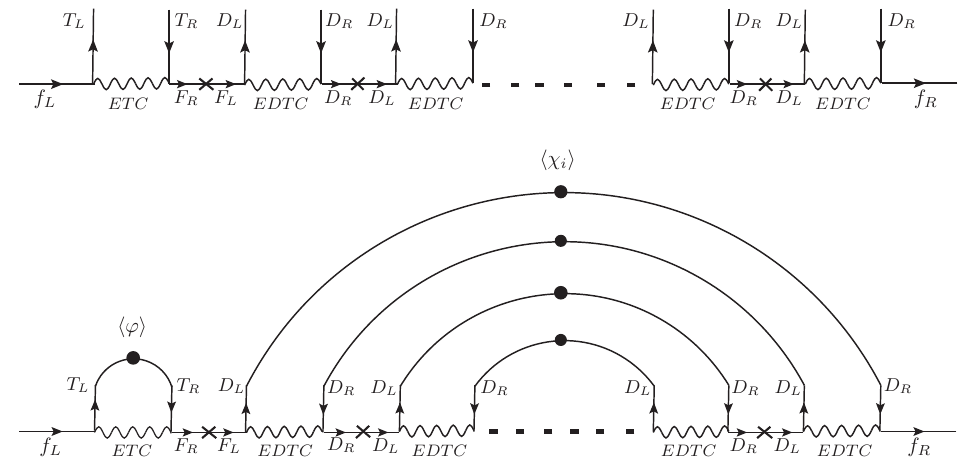}
    \caption{The Feynman diagrams for the masses of the  charged fermions in the  dark-technicolor model.}
 \label{fig1}	
 \end{figure}

We further assume that the TC fermions,  the left-handed SM fermions, and  the $F_R$ fermions are accommodated  in an extended TC (ETC) symmetry, and the DTC fermions,  the right-handed SM singlet fermions, and the $F_L$ fermions unify in a different extended DTC (EDTC) symmetry.   This assumed setup leads to the required interactions for generating the charged fermion masses and mixing. For instance, creation of  masses and mixing of the charged fermions is depicted in a generic Feynman diagram in figure \ref{fig1}. On the top, we show the generic interactions of the TC, DTC, and F fermions, and in the bottom, the formation of different multi-fermion condensates playing the role of the field $\chi_r$,  is shown  \cite{Abbas:2025ser}.

We can write the  multi-fermion condensate as  \cite{Aoki:1983yy}, 
\be 
\label{VEV_h}
\langle  ( \bar{\psi}_L \psi_R )^n \rangle \sim \left(  \Lambda \exp(k \Delta \chi) \right)^{3n},
\ee
where $\Delta \chi$ shows the chirality of the corresponding operator, $k$ denotes a constant, and $\Lambda$ is  the scale of the underlying gauge dynamics.

The masses of the charged fermions are now given by,
\bea
\label{TC_masses}
m_{f} & = & y_f  \frac{\Lambda_{\text{TC}}^{3}}{\Lambda_{\text{ETC}}^2}  \dfrac{1}{\Lambda_{\text{F}}} \frac{\Lambda_{\text{DTC}}^{n_i + 1}}{\Lambda_{\text{EDTC}}^{n_i}} \exp(n_i k),~
\eea
where $n_i = 2,4, \cdots 2 n $ shows the number of fermions in a multi-fermion chiral condensate playing the role of the VEV $ \langle \chi_r \rangle$ in the SHVM.

Comparing with equation \ref{TC_masses}, we can write,
\bea
\epsilon_r \propto \dfrac{1}{\Lambda_{\text{F}}} \frac{\Lambda_{\text{DTC}}^{n_i + 1}}{\Lambda_{\text{EDTC}}^{n_i}} \exp(n_i k).
\eea

\begin{figure}[H]
	\centering
 \includegraphics[width=\linewidth]{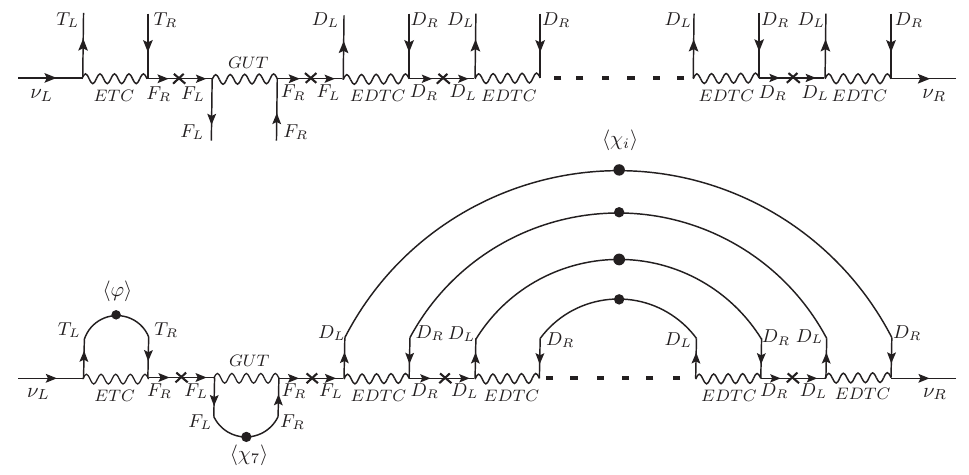}
    \caption{The Feynman diagrams leading to  dimension-6 operators, which are responsible for neutrino masses, are given in equation \ref{mass_Nu}.    On the top, we show the generic interactions involving the SM, TC, DTC gauge sectors mediated by ETC, EDTC, and GUT gauge bosons.  At the bottom, the formation of the fermionic condensate $\langle \bar{F}_L F_R \rangle$, which is the VEV $\langle \chi_7 \rangle$ in the SHVM, is depicted. }
 \label{fig2}	
 \end{figure}

Neutrino masses are generated by assuming that ETC and EDTC symmetries are eventually accommodated  in a GUT theory.  This leads to the creation of the dimension-6 operators responsible for neutrino masses,  given in equation \ref{mass_Nu}. The corresponding interactions are depicted in  figure \ref{fig2} where the  GUT gauge bosons mediate the  interactions  between the $F_L$ and $F_R$ fermions. The role of the VEV $\langle \chi_7 \rangle$ is performed by the chiral condensate  $\langle \bar{F}_L F_R \rangle$.

 The neutrinos masses can be written as,
\bea
\label{TC_nmasses}
m_{\nu} & \approx & y_f  \frac{\Lambda_{\text{TC}}^{3}}{\Lambda_{\text{ETC}}^2}  \dfrac{1}{\Lambda_{\text{F}}} \frac{\Lambda_{\text{DTC}}^{n_i + 1}}{\Lambda_{\text{EDTC}}^{n_i}} \exp(n_i k)  \dfrac{1}{\Lambda_{\text{F}}} \frac{\Lambda_{\text{F}}^{3}}{\Lambda_{\text{GUT}}^{2}} \exp(2 k),
\eea
where,  
\bea
\epsilon_{7} \propto  \dfrac{1}{\Lambda_{\text{F}}} \frac{\Lambda_{\text{F}}^{3}}{\Lambda_{\text{GUT}}^{2}} \exp(2 k).
\eea
For more  phenomenological details,  see ref. \cite{Abbas:2025ser}.

\end{document}